\documentclass[11pt,a4paper]{article}
\usepackage{jheppub}

\usepackage{multirow, graphicx,amssymb,url,mathrsfs,amsmath}
\usepackage{wrapfig,boxedminipage,setspace,subfigure,epsfig}
\usepackage{amsxtra,amstext,latexsym,dsfont,amsfonts}
\usepackage{color,eucal}
\usepackage[dvipsnames]{xcolor}
\usepackage{float}
\usepackage{slashed,comment}
\usepackage{kotex}
\usepackage{colortbl}
\usepackage{multirow}
\usepackage{tikz}
\usepackage{amsmath}
\usetikzlibrary{tikzmark}
\usetikzlibrary{arrows}

\usepackage{tabularx, array}
\newcolumntype{L}[1]{>{\raggedright\arraybackslash}p{#1}}
\newcolumntype{C}[1]{>{\centering\arraybackslash}p{#1}}
\newcolumntype{R}[1]{>{\raggedleft\arraybackslash}p{#1}}

\usepackage{pdflscape}

\title{\huge Mechanical stability of homogeneous holographic solids under finite shear strain}

\author[a,b]{Matteo Baggioli,}
\author[c,d,e]{Li Li,}
\author[f]{Wei-Jia Li,}
\author[c,d]{and Hao-Tian Sun}

\emailAdd{b.matteo@sjtu.edu.cn}
\emailAdd{liliphy@itp.ac.cn}
\emailAdd{weijiali@dlut.edu.cn}
\emailAdd{sunhaotian@itp.ac.cn}

\affiliation[a]{Wilczek Quantum Center, School of Physics and Astronomy, Shanghai Jiao Tong University, Shanghai 200240, China}
\affiliation[b]{Shanghai Research Center for Quantum Sciences, Shanghai 201315, China}
\affiliation[c]{CAS Key Laboratory of Theoretical Physics, Institute of Theoretical Physics,
Chinese Academy of Sciences, Beijing 100190, China}

\affiliation[d]{School of Physical Sciences, University of Chinese Academy of Sciences,\\
No.19A Yuquan Road, Beijing 100049, China}

\affiliation[e]{School of Fundamental Physics and Mathematical Sciences, Hangzhou Institute for Advanced Study, UCAS, Hangzhou 310024, China}
\affiliation[f]{Institute of Theoretical Physics, School of Physics, Dalian University of Technology,
Dalian 116024, China}

\abstract{We study the linear stability of holographic homogeneous solids (HHS) at finite temperature and in presence of a background shear strain by means of a large scale quasi-normal mode analysis which extends beyond the hydrodynamic limit. We find that mechanical instability can arise either as a result of a complex speed of sound -- gradient instability -- or of a negative diffusion constant. Surprisingly, the simplest HHS models are linearly stable for arbitrarily large values of the background strain. For more complex HHS, the onset of the diffusive instability always precedes that of the gradient instability, which becomes the dominant destabilizing process only above a critical value of the background shear strain. Finally, we observe that the critical strains for the two instabilities approach each other at low temperatures. We conclude by presenting a phase diagram for HHS as a function of temperature and background shear strain which shows interesting similarities with the physics of superfluids in presence of background superfluid velocity.}

\begin{document}
\maketitle

\section{Introduction}
A crystalline solid is a phase of matter characterized by translational long-range order: the spontaneous breaking of translational invariance, possibly down to a discrete sub-group~\cite{Lubensky}. Holographic solids~\cite{PhysRevLett.120.171602,RevModPhys.95.011001} are not different in this perspective, as they are also built upon the same symmetry breaking pattern. The long-range order endows solids with their rigidity and with their ability to respond elastically to mechanical deformations\footnote{To be precise, this statement is correct only for crystalline ordered solids. In amorphous systems, the emergent rigidity and the associated elastic response have a more complex, and still not well understood, origin.}. Phonons, or sound-waves, are the corresponding emergent Goldstone modes~\cite{leutwyler1996phonons}. Despite the behavior of crystalline solids under infinitesimal mechanical deformations is well described by linear elasticity theory \cite{landau7}, their mechanical response beyond the linear regime, \textit{i.e.}, for finite deformations, is more difficult to be rationalized. Nonlinear elasticity \cite{ZAMM:ZAMM19850650903} is then usually constructed in terms of empirical models based on phenomenological assumptions and a unifying picture is still missing. These problems become even more severe in presence of finite temperature and dissipative effects, which render the response of the material viscoelastic rather than purely elastic \cite{HYUN20111697}. On top of that, irreversible plastic deformations are inevitable for sufficiently large deformations, and they are indeed fundamental to describe the failure of solid materials. Once again, apart from phenomenological elasto-plastic models \cite{RevModPhys.90.045006}, a complete understanding of the non-linear behavior of complex solids has not been achieved yet.
\begin{figure}
    \centering
    \includegraphics[width=0.55\linewidth]{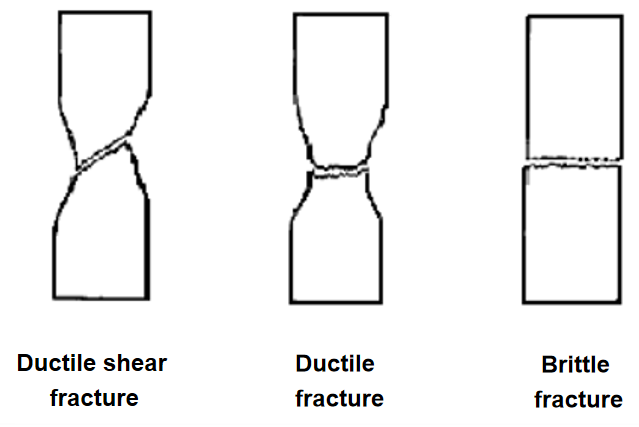}
    \caption{A visual representation of different types of mechanical failure in solid materials.}
    \label{fig:0}
\end{figure}

From a material science and engineering perspective, understanding the mechanical failure of solid materials is a question of primary importance. The main concern of material failure theory \cite{10.1093/acprof:oso/9780199662111.001.0001} is to predict under which conditions a solid fails under external loads. Failure is necessarily connected to some sort of instability of the system and it is inevitable for large enough deformations. The instability, and therefore the failure, could be either microscopic or macroscopic. We define an instability macroscopic if it can be derived from the macroscopic effective equations of the medium, \textit{i.e.}, when it concerns the collective long-wavelength dynamics. A macroscopic failure is global and it could in principle be derived from simple arguments related to thermodynamics, energetic considerations or even nonlinear elasticity theory. On the other hand, a microscopic instability, usually associated to crack initiation and propagation, pertains to the structure of the material at short scales, and could be somehow invisible within the homogeneous effective description.

In addition to the above classification, which is based on the length-scale of interest, mechanical failure can be also divided into \textit{brittle} and \textit{ductile} (see Fig.~\ref{fig:0} for an illustration). Brittle failure is usually accompanied by an extremely rapid propagation of cracks, so quickly that no plastic deformation has time to take place. Brittle failure causes a catastrophic fracture which destroys the structural integrity of the whole material. On the other side, ductile materials deform plastically and they have the ability to support more stress, slowing the fracture process and avoiding a catastrophic failure. Their failure is anticipated by a necking instability in which the cross-sectional area of the sampled diminishes when undergoing deformation. Necking and ductility are also often associated to yielding, the breakdown of the elastic response in a solid and the onset of plastic deformation.

\begin{figure}
    \centering
    \includegraphics[width=0.50\linewidth]{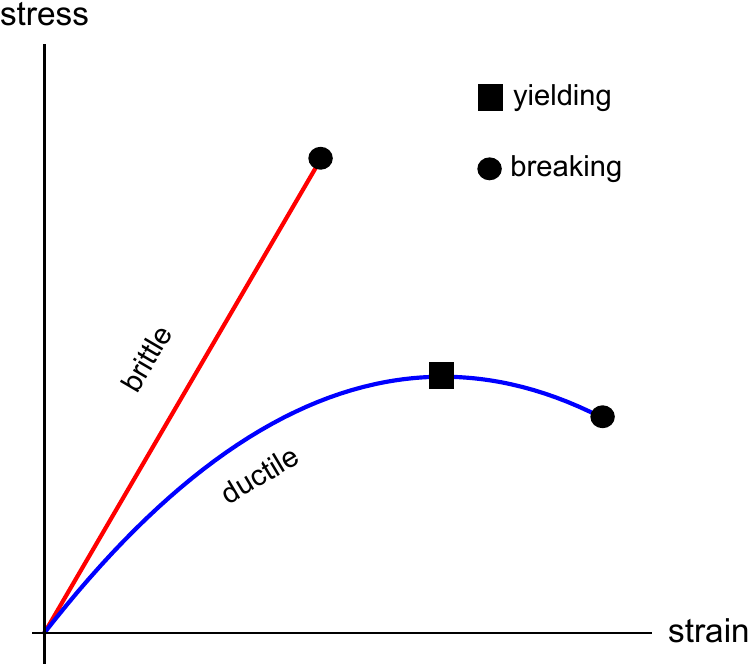}
    \caption{A cartoon of the stress-strain curves for ductile and brittle materials. The black dots indicate the breaking strength and the black square the yielding point, defined as the maximum of the stress-strain curve.}
    \label{fig:1}
\end{figure}

All in all, whether a material displays a ductile or brittle behavior, and which is the stress/strain at which the failure appears, depend on several factors related to the composition of the material and to the external conditions as well (\textit{e.g.}, temperature). A long list of criteria have been proposed in the literature, but it is hard to encompass such a vast and complex phenomenology under a few simple assumptions. Nevertheless, a perhaps naive differentiation of the two phenomena can be anticipated looking at the behavior of the stress-strain curve for finite deformations (see Fig.~\ref{fig:1}). Brittle materials are characterized by a quick increase of the stress which suddenly ends at the breaking point. Ductile materials, on the contrary, are able to support larger strains. The induced stress diminishes with increasing the strain and the stress-strain curve bends down approaching a plateau-like form. The yielding point usually appears right before that regime and it marks the end of the elastic response. After the yielding point, the response is purely plastic and the stress becomes almost independent of the strain up to the breaking point, which in ductile materials is located at much larger values of the strain than in brittle ones.

Holographic models with spontaneously broken translations, \textit{i.e.}, holographic solids, have attracted great interest in the last years \cite{RevModPhys.95.011001}, as a possible novel route to tackle the notoriously hard problem of strange metals and spontaneous orders in strongly correlated phases of matter. A particularly simple sub-class of models is given by the so-called ``homogeneous holographic solids'', in which translational symmetry is spontaneously broken while retaining homogeneity, owing to the existence of an internal global symmetry. Among the various possibilities, the holographic solid axion model \cite{PhysRevLett.114.251602,Alberte:2015isw}, which was found as an extension of the same model for momentum dissipation \cite{Andrade:2013gsa} (see \cite{Baggioli:2021xuv} for a review), is the only one which provides an analytical solution, and therefore allows for extensive computations. The holographic model displays propagating phonons as dictated by elasticity theory \cite{PhysRevLett.120.171602}, and its low-energy effective description is in perfect agreement with viscoelasticity \cite{Armas:2019sbe,Armas:2020bmo}, both at zero and finite charge density \cite{Ammon:2019apj,Ammon:2020xyv,Baggioli:2020edn}. The linear viscoelastic dynamics \cite{Andrade:2019zey,Baggioli:2019elg} and nonlinear elastic response \cite{Baggioli:2019mck,Baggioli:2020qdg,Pan:2021cux} of this holographic model have been studied in detail in the past.

In this work, our task is to make a step further and perform a failure analysis of these holographic solids under large static shear deformations. From a more technical perspective, this amounts to investigate the instabilities of these holographic models upon applying a background external static shear strain. In \cite{Baggioli:2020qdg,Alberte:2018doe}, some general criteria based on the stress-strain curves and the dispersion relations in the decoupling limit have been provided. Nevertheless, a full-fledged analysis is still missing. Here, we study the dynamical stability of the model based on a quasi-normal mode analysis in presence of large background shear strain. In other words, we linearly perturb our system around a state with a large shear strain and check its linearized dynamics. Since the background shear strain breaks rotational symmetry, the dispersion relation of the low-energy modes will depend on the angle of propagation. In a way, this is similar to the case of a holographic superfluid model with superfluid velocity \cite{Amado:2013aea}, where the latter plays the role of our external shear strain\footnote{In this analogy, the failure of the solid corresponds to the instability of the superfluid phase, \textit{i.e.}, the Landau criterion.}.

The concept of ideal strength of a material, which defines the onset of mechanical instability, is indeed closely related to the dispersion of the phonon modes in the solid, and possible phonon instabilities. Those happen whenever the energy of one mode becomes negative, or more in general when the imaginary part of its frequency $\omega(k)$ becomes positive. In the case of an ``\textit{elastic instability}'', the instability appears for small values of the wave-vector $k$, and can be in principle detected by inspecting the elastic moduli of the system or looking at its hydrodynamic (or even thermodynamic) description. Nevertheless, instabilities can appear for arbitrary, and even large, values of the wave-vector, and in that sense being more microscopic (\textit{e.g.}, holographic charge density waves \cite{Nakamura:2009tf,Donos:2011qt,Donos:2013gda,Donos:2013wia} or pair density waves \cite{Cai:2017qdz,Cremonini:2017usb,Cremonini:2016rbd}). Phonon instabilities under mechanical deformations have been explored in several materials (\textit{e.g.}, \cite{PhysRevLett.91.135501,RevModPhys.84.945,PhysRevLett.105.245502,PhysRevB.85.235407,PhysRevB.76.064120,PhysRevB.89.184111,Wang_2016}). Interestingly, these elastic instabilities often appear before the peak in the stress-strain curve \cite{PhysRevLett.91.135501}, at which the elastic response is completely destroyed into the plastic flow regime. Also, these instabilities are frequently related to soft modes and structural phase transitions \cite{PhysRevB.13.4877,PhysRevB.94.014109,PhysRevB.86.054118,Venkataraman1979}, the classical example being the Kohn anomaly precursor of the formation of charge density waves \cite{PhysRev.126.1693}.

In the rest of this work, we will observe that holographic homogeneous solids display a complex structure of instabilities which can be nevertheless classified into two types:
\begin{enumerate}
    \item \textbf{Sound instability} (or gradient instability). This instability manifests itself in a speed of sound which becomes complex, \textit{i.e.}, $v^2<0$.
    \item \textbf{Diffusive instability}. This second type corresponds to a diffusion constant which becomes negative, \textit{i.e.}, $D<0$.
\end{enumerate}
In both cases, the instability occurs when the imaginary part of the frequency of a certain mode becomes positive, destabilizing the initial background solution. As we will see, the diffusive instability will always be the first to appear (\textit{i.e.}, the one with the lowest critical strain) in the holographic models considered in this work. In other words, the sound instability will appear only beyond a specific threshold for the background shear strain.

As a general remark, the mechanical failure of a solid material involves the creation, and
 proliferation of inhomogeneous structures such as defects, cracks and voids. Mechanical failure is often the result of complex microscopic inhomogeneous processes. In an effective description, that is agnostic of the microscopic physical processes, and it is natural to assume that those inhomogeinities correspond to inhomogeneous instabilities as the ones just mentioned.

The manuscript is organized as follows. In Section \ref{sec1}, we describe the homogeneous holographic solid models considered in this manuscript and briefly recap their viscoelastic linear dynamics and their static nonlinear elastic properties. In Section \ref{sec2}, we study the mechanical stability of the simplest of these models under background finite shear strain. In Section \ref{sec3}, we generalize our analysis to a larger class of models and display a complex structure of instabilities. We end up this section by drawing a phase diagram for the dual solid system. Finally, in section \ref{sec4}, we summarize our results and conclude with a few comments for the future and some analogies with the instabilities appearing in superfluids with finite superfluid velocity.
\section{Holographic setup and its viscoelastic description}\label{sec1}
\subsection{Axion model and homogeneous solids}
We start with the most general 4-dimensional Einstein-axion action \cite{Alberte:2015isw},
\begin{equation}\label{ac}
    S=\int d^4x \sqrt{-g}\,\left[R-2\Lambda - m^2\, V(X,Z)\right],
\end{equation}
with
\begin{equation}
\mathcal{X}_{I J} \equiv \partial_{\mu} \phi_I \partial^{\mu} \phi_J,\quad X \equiv \frac{1}{2} \operatorname{Tr}\left(\mathcal{X}_{I J}\right),\quad Z \equiv \operatorname{det}\left(\mathcal{X}_{I J}\right), \quad I=1,2\,.   
\end{equation}
Here $R$ is the Ricci scalar, $m^2$ is a parameter related to the graviton mass, $V$ is an arbitrary scalar function and $\Lambda$ is the cosmological constant. The action in Eq.~\eqref{ac} is invariant under global internal shifts $\phi^I\rightarrow \phi^I+c^I$, since it only involves derivative terms. The general equations of motion are given by
\begin{equation} \label{eom1}
\begin{split}
&\nabla_{\mu}\left(V_{X} \nabla^{\mu} \phi^1+2 V_{Z}\nabla^{\mu}\phi^1 \mathcal{X}^{22}-2 V_{Z} \nabla^{\mu} \phi^{2}\mathcal{X}^{12}\right)=0\,, \\
&\nabla_{\mu}\left(V_{X} \nabla^{\mu} \phi^2+2 V_{Z}\nabla^{\mu}\phi^2 \mathcal{X}^{11}-2 V_{Z} \nabla^{\mu} \phi^{1}\mathcal{X}^{21}\right)=0\,, \\
&R_{\mu \nu}-\frac{1}{2}R g_{\mu \nu}+\Lambda g_{\mu \nu}=m^2\mathcal{T}_{\mu\nu}\,, 
\end{split}
\end{equation}
with $\mathcal{T}_{\mu\nu}$ the energy-momentum tensor
\begin{align}
&\mathcal{T}_{\mu\nu}=-\frac{1}{2} g_{\mu \nu}V(X,Z)+\frac{V_X}{2} \sum_{I=1}^{2}\partial_{\mu} \phi^{I} \partial_{\nu} \phi^{I}\notag\\
&+V_Z (\mathcal{X}^{22}\partial_{\mu} \phi^{1} \partial_{\nu} \phi^{1}+\mathcal{X}^{11}\partial_{\mu} \phi^{2} \partial_{\nu} \phi^{2}-\mathcal{X}^{12}\partial_{\mu} \phi^{2} \partial_{\nu} \phi^{1}-\mathcal{X}^{21}\partial_{\mu} \phi^{1} \partial_{\nu} \phi^{2})\,, \label{eom2}    
\end{align}
and
\begin{align}
&V_Z=\frac{\partial V(X,Z)}{\partial Z},\quad V_X=\frac{\partial V(X,Z)}{\partial X}\,.
\end{align}

Remarkably, many previous studies \cite{Baggioli:2021xuv} have shown that this large class of holographic models allows to break spatial translations while retaining the homogeneity of the dual boundary system as long as the bulk profile of the axions is chosen to be
\begin{equation}\label{profile}
\phi^I\,=\,{M^I}_i\,x^i,
\end{equation}
where $M^I_i$ is a $2\times2$ matrix whose physical meaning will appear clear later. Previous studies made the simplest choice
\begin{equation}\label{profileiso}
{M^I}_i\,=\,\alpha\,{\delta^I}_i, 
\end{equation}
which fully preserves the $\text{SO}(2)$ rotation in the $x$-$y$ plane. Within the isotropic setup, the background metric of the black hole takes the following form,
\begin{equation}\label{bgmetric}
ds^2=\frac{L^2}{u^2}\left[-f(u)dt^2+\frac{1}{f(u)}du^2+dx^2+dy^2\right],
\end{equation}
where the square of the AdS radius $L^2=-3/\Lambda$. Here, $u$ is the radial coordinate ranging from the AdS boundary, $u=0$, to the black hole horizon, $u=u_h$. The blackening factor $f(u)$ can be easily calculated by inserting \eqref{bgmetric} into the the Einstein equations, and it is given by 
\begin{equation}
f(u)=u^3\int^{u_h}_u ds\left(\frac{3}{s^4}-\frac {m^2\,V(\bar{X},\bar{Z})}{s^4}\right)ds,
\end{equation}
where the ``bar" denotes the background value of the respective functions. For the simplest case~\eqref{profileiso}, we have $\bar{X}=\alpha^2 u^2$ and $\bar{Z}=\alpha^4 u^4$.
Furthermore, the Hawking temperature is given by
\begin{equation}
T=-\frac{f'(u_h)}{4\pi}=\frac{3\,-m^2\,V(\bar{X}_h,\bar{Z}_h)}{4\pi u_h},
\end{equation}
and the entropy density reads
\begin{equation}
s=\frac{2\pi}{u_h^2}.
\end{equation}
Considering the case of a monotonic power law function\footnote{It is sufficient to have this power-law behavior close to the boundary $u\rightarrow 0$ and not in the entire bulk.}, $V(X,Z)=X^NZ^M$, the asymptotic behavior of the axion fields near the boundary takes the general form
\begin{equation}
\phi^I(u,\Vec{x},t)=\phi_{(0)}^I(\Vec{x},t)\,\dots\,+\,\phi_{(1)}^I(\Vec{x},t)\,u^{5-2N-4M}\,+\,\dots,
\end{equation}
where $(\Vec{x},t)$ denote the space-time coordinates on the boundary. In the standard quantization scheme, the coefficient of the leading order term in the expansion is interpreted as the external source for the boundary operator $\mathcal{O}^I$, while the other corresponds to its expectation value, $\langle \mathcal{O}^I\rangle$. Therefore, by tuning the exponents, $N$ and $M$, we can achieve different scenarios \cite{PhysRevLett.120.171602} :
\begin{itemize}
\item If $5-2N-4M>0$, $\phi_{(0)}^I(x^\mu)$ is the leading term in the above expansion and the profile \eqref{profile} corresponds to source the dual field theory with a space-dependent coupling which breaks translations explicitly. This leads to momentum dissipation in the dual field theory \cite{Andrade:2013gsa,Davison:2013jba}.
\item If $5-2N-4M<0$, $\phi_{(0)}^I(x^\mu)$ is now the subleading term in the above expansion and the same profile \eqref{profile} should be interpreted as a space-dependent expectation value $\langle \mathcal{O}^I\rangle \propto x^i$, in absence of any source, which corresponds to the onset of spontaneous symmetry breaking (SSB) in the boundary field theory \cite{PhysRevLett.120.171602}. This second route is analogous to introducing a boundary kinetic term for the scalar fields \cite{Armas:2019sbe}, and therefore dynamical elastic interactions in the boundary field theory. In this sense, it is very similar to the procedure to introduce dynamical electromagnetism and Coulomb interactions in holography using mixed boundary conditions (see, \textit{e.g.}, \cite{Ahn:2022azl,Baggioli:2023oxa}).
\end{itemize} 
In the rest of this paper, we will focus on the later case.  {The holographic renormalization of axion models with spontaneously broken translations was already considered in previous studies. All the models corresponding to the SSB case (in the standard quantization) are less UV relevant than the standard linear axion model. Therefore, no extra counterterms are needed. For more details about the holographic renormalization for this type of models, please refer to the appendices in~\cite{Ammon:2020xyv,Ji:2022ovs}.}

\subsection{Viscoelasticity and linear response}
In order to compute the linear response of the system upon external deformations, we introduce the fluctuating fields 
\begin{equation}\label{isolinear}
\delta \chi_A (u,t,\Vec{x})=\int^{+\infty}_{-\infty}\frac{d\omega \,d^2k}{(2\pi)^3}\, e^{i\,\left(\Vec{k} \cdot \Vec{x}-\omega\,t\right)}\, \delta \chi_A (u,\omega,\Vec{k})
\end{equation}
on top of the background solution described in the previous section. Here, $A$ is just a collective label. We remind the reader that the boundary has two spatial dimensions, and therefore $\Vec{k}\equiv (k_x,k_y)$.
Because of isotropy, one can always divide the complete set of the fluctuating modes into longitudinal and transverse sectors which include respectively,
\begin{align}
  &\delta \chi_L:=\{\delta g_{tt},\,\delta g_{xx},\,\delta g_{yy},\,\delta g_{tu},\,\delta g_{uu},\,\delta \phi^{x}\},\\ 
&\delta \chi_T:=\{\delta g_{ty},\,\delta g_{xy},\,\delta g_{uy},\,\delta \phi^{y}\}.
\end{align}
Without loss of generality, the wave vector $\Vec{k}$ has been taken along the $x$-direction, $\vec{k}=(k,0)$. Then, the dispersion relations (which are poles of retarded correlators) of long-lived excitations on boundary can be extracted from the low-lying quasi-normal modes of the black hole in the bulk.

In linear response theory, and in absence of finite background shear deformations, the homogeneous solids (in absence of finite charge density) exhibit the following hydrodynamic modes:
\begin{align}
    & \text{transverse sector:}\quad\; \omega\,=\,\pm\,v_T\,k\,-\,\frac{i}{2}\,\Gamma_T\,k^2\,,\label{tmode}\\
    &\text{longitudinal sector:}\;\;\omega\,=\,\pm\,v_L\,k\,-\,\frac{i}{2}\,\Gamma_L\,k^2\,,\quad \omega\,=\,-\,i\,D_\phi\,k^2\,,\label{lmode}
\end{align}
where $k$ is the magnitude of the wave vector $\Vec{k}$. In the transverse sector, one obtains a pair of propagating shear sound modes with speed $v_T$ and attenuation constant $\Gamma_T$. In the longitudinal sector, one has a pair of longitudinal propagating sound modes with speed $v_L$ and attenuation constant $\Gamma_L$ and a diffusive mode with diffusion constant $D_\phi$. For more details, see ~\cite{Armas:2019sbe,Ammon:2020xyv,Armas:2020bmo}.

For a relativistic neutral conformal solid, the various transport coefficients appearing in the dispersion relations are given by
~\cite{Armas:2019sbe,Ammon:2020xyv}:
\begin{gather}
    v_T^2
    = \frac{G}{\chi_{\pi\pi}}\,,\qquad 
    v_L^2 
    =\frac{1}{2}+ v_T^2\,,
    \quad 
  \Gamma_T
  = \frac{\eta}{\chi_{\pi\pi}}
  +\frac{G}{\sigma}\frac{s^2 T^2}{\chi_{\pi\pi}^2}\,, \nonumber\\
 \Gamma_L
  = \frac{\eta}{\chi_{\pi\pi}}
  +\frac{T^2s^2 G^2}{\sigma\chi_{\pi\pi}^3v_L^2}\,, \quad
  D_\phi
  = \frac{Ts^2/\sigma}{s+\partial_T \mathcal{P}}
  \frac{B+G-\mathcal{P}}{\chi_{\pi\pi} + 2G}\,.
  \label{eq:modes}
\end{gather}
The relation between the transverse and longitudinal sound speeds is a result of conformal invariance \cite{Esposito:2017qpj}.
Here, $\chi_{\pi\pi}= \varepsilon + p +\mathcal{P}$ is the momentum susceptibility and other coefficients are derived in terms of the following thermodynamic relations and Kubo formulas,
\begin{gather}
    \varepsilon = \langle T^{tt} \rangle\,, \quad
    p = -\Omega\,, \quad
    \mathcal{P} = \langle T^{xx} \rangle + \Omega\,,\quad
    \chi_{\pi\pi}v_L^2 = \lim_{\omega\to0}\lim_{k\to0}
    \mathrm{Re}\,G^R_{T^{xx}T^{xx}}\,,
    \nonumber\\
    G = \lim_{\omega\to0}\lim_{k\to0}
    \mathrm{Re}\,G^R_{T^{xy}T^{xy}}\,,
    \quad 
    \eta = -\lim_{\omega\to0}\lim_{k\to0}
    \frac{1}{\omega}\mathrm{Im}\,G^R_{T^{xy}T^{xy}}\,,
    \nonumber\\
    B = (3\mathcal{P}-T\partial_T\mathcal{P})/2\,,\quad\quad
    \frac{(\varepsilon+p)^2}{\sigma\chi_{\pi\pi}^2} 
    = \lim_{\omega\to0}\lim_{k\to0}   \omega\,\mathrm{Im}\,G^R_{\Phi^{x}\Phi^{x}}\,,
    \label{kubos}
\end{gather}
where $\Omega$ is the free energy density, $T^{\mu\nu}$ the boundary stress-energy tensor and $\Phi^i$ the Goldstone operator associated with the spontaneously translation symmetry breaking. Notice that the shear modulus $G$\footnote{{It was found that, in the axion model, there is no well-defined melting temperature, intended as a first-order thermodynamic phase transition between a liquid phase and a solid phase. Instread, the shear modulus just decreases smoothly to zero as $T/\alpha \rightarrow \infty$ \cite{PhysRevLett.120.171602}.}}, the shear viscosity $\eta$, and the dissipative parameter $\sigma$ can be obtained from the Kubo formulas. Finally, the lattice pressure $\mathcal{P}$ quantifies the difference between the thermodynamic and mechanical pressures and represents an additional contribution to the mechanical pressure as a result of working around a state which does not minimize the free energy \cite{Donos:2013cka,RevModPhys.95.011001}. 

\begin{figure}
    \centering
 \includegraphics[width=0.48\textwidth]{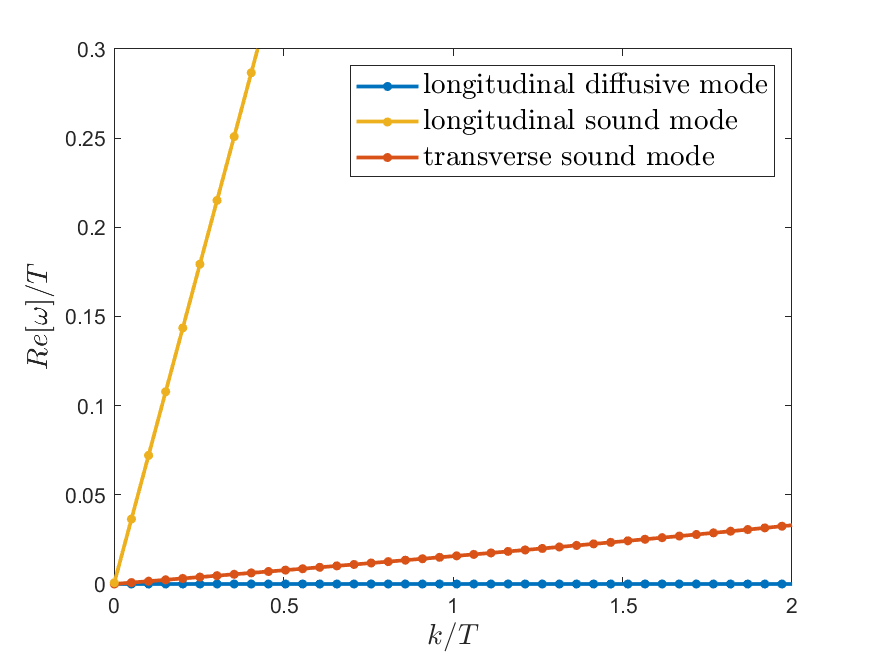}
    \includegraphics[width=0.48\textwidth]{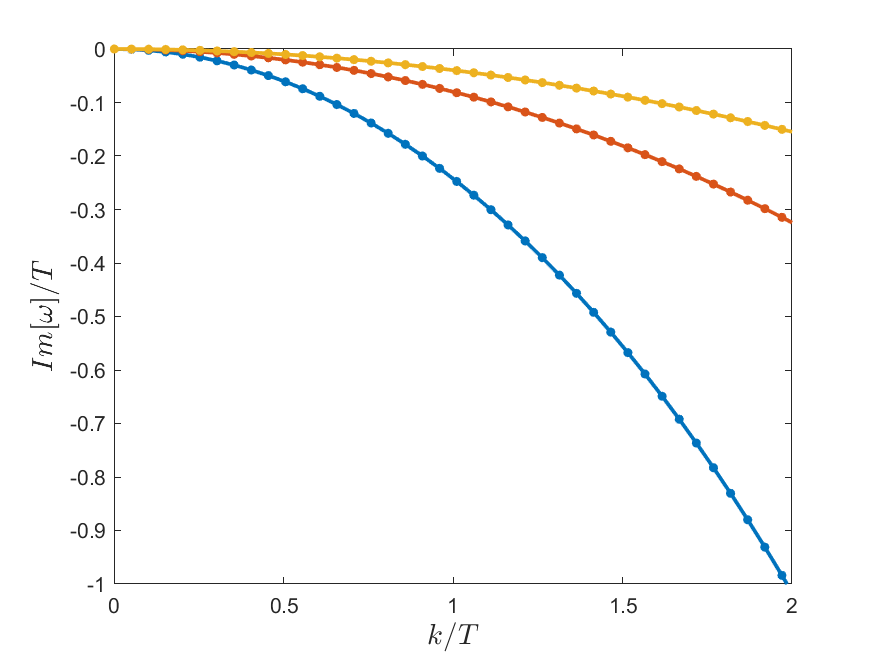}
    \caption{Dispersion relation of the lowest excitations at zero background strain $\epsilon=0$. The rest of the parameters are fixed to $m L= 1$ and $T/\alpha=0.8$.  The bullets are the numerical data and the solid lines the predictions from hydrodynamics of Eq.~\eqref{eq:modes}.}
    \label{fig:3}
\end{figure}

In Fig.~\ref{fig:3}, we show a concrete example of the dispersion relations of the hydrodynamic modes for the simplest potential which implements the SSB of translations, $V(X,Z)=X^3$.\footnote{In the simple case of monomial potentials, only one of the dimensionful parameters, $m$ or $\alpha$, is necessary to characterize the strength of the spontaneous breaking of translation. In the rest of this paper, we always fix $m=L=1$ and treat $\alpha$ as a free parameter.} We observe a pair of sound modes in the transverse channel, a pair of sound modes with larger speed and a single diffusive mode in the longitudinal channel. Furthermore, one can verify quantitatively that, at long distances (small $k$), the dispersion relations are in perfect agreement with the analytic relations \eqref{tmode}-\eqref{kubos} shown with solid lines in Fig.~\ref{fig:3}.

\subsection{Beyond linear response}

Whenever the external strain applied to an ideal solid (with no dissipation) is infinitesimal, linear elastic response theory implies a simple relation between the strain $\mathcal{E}_{kl}$ and the induced stress $\sigma_{ij}$
\begin{align}\label{linear}
\sigma_{ij}\,=\,C_{ijkl}\,\mathcal{E}_{kl}\, +\,\dots,
\end{align}
where $C_{ijkl}$ is the elastic tensor, determined by the elastic constants of the material. However, when the system undertakes a finite, and eventually large, strain, a richer behavior emerges and the stress-strain relation is in general no longer linear,
\begin{equation}
    \sigma_{ij}=\Sigma(\mathcal{E}_{kl}).
\end{equation}
In this case, the system enters into the nonlinear elastic regime, where the perturbative linear approximation in Eq.~\eqref{linear} breaks down. Recently, the nonlinear response within the holographic axion model has been studied \cite{Baggioli:2020qdg} (see also \cite{Pan:2021cux,Baggioli:2022aft,Ji:2022ovs}). 

In order to introduce a finite and static background shear strain in the holographic model, one can choose a more general profile for the bulk axion fields,
\begin{align}\label{profileaniso}
M^I_i\,=\, \alpha\,\left(\begin{array}{cc}
\sqrt{1+\epsilon^2/4} & \epsilon/2 \\
\epsilon/2 & \sqrt{1+\epsilon^2/4}
\end{array}\right)\,.
\end{align}
From the perspective of effective field theory \cite{Alberte:2018doe}, $\alpha$ characterizes the variation of the volume (or area in two spatial dimensions) of the system, hence corresponding to a background bulk strain. On the other hand, $\epsilon$ corresponds to an anisotropic mechanical deformation which preserves volume, but not angles, \textit{i.e.}, an external shear strain.

Because of the introduction of off-diagonal terms in \eqref{profileaniso}, the background solution has to be modified into the more general form
\begin{align}\label{anisobgmetric}
ds^2=\frac{1}{u^2}\left[-f(u)e^{-\chi(u)}dt^2+\frac{1}{f(u)}du^2+\gamma_{ij}(u)dx^idx^j\right],
\end{align}
 where $\gamma_{ij}(u)=\delta_{ij}$ in the absence of shear deformation. 
For later convenience, we further introduce the following notations,
\begin{align}\label{gammanot}
\gamma=\left(\begin{array}{cc}
\text{cosh}\,h(u) & \text{sinh}\,h(u) \\
\text{sinh}\,h(u) & \text{cosh}\,h(u)
\end{array}\right)\,
\end{align}
and 
\begin{align}\label{profileaniso2}
M^I_i=\alpha\,\left(\begin{array}{cc}
\text{cosh}\,(\Omega/2) & \text{sinh}\,(\Omega/2) \\
\text{sinh}\,(\Omega/2) & \text{cosh}\,h(\Omega/2)
\end{array}\right)\,
\end{align}
with the unstrained reference configuration corresponding therefore to $\Omega=0$. Then, the background shear strain can be expressed as
\begin{align}\label{strain}
\epsilon\,=\,2\,\text{sinh}\left(\frac{\Omega}{2}\right).
\end{align}
We refer to \cite{Baggioli:2022aft} for a detailed discussion about the meaning and the interpretation of the holographic setup in comparison to standard non-linear elasticity theory.\footnote{{One can also choose a base that diagonalizes the matrix $M^I_i$ in Eq.~\eqref{profileaniso} as well as the spatial components of the spatial metric $\gamma_{ij}$ in Eq.~\eqref{gammanot}. This new coordinate system is nothing but the one aligned with the principle axis of the shear. This alternative choice is equivalent to the one presented in this work.} }

For the above ansatz, the background equations of motion read
\begin{align}\label{bgeom}
2\,\chi'\,-\,u\,{h'}^2=0\,,\\
f\,(u\,\chi'\,+\,6\,)\,+\,\left[m^2\,V\,(\bar{X},\bar{Z})\,-\,2\,u\,f'\,-\,6\right]\,=\,0\,,\\
h''\,+h'\,\left({-}\,\frac{f'}{f}\,-\,\frac{2}{u}\right)-\frac{1}{4}\,u\,{h'}^3\,-\,\frac{\alpha^2\,\text{sinh}\,(\Omega-h)\,m^2 V_X(\bar{X},\bar{Z})}{f}=0,
\end{align}
where the background values of $X$ and $Z$ are given by $\bar{X}\,=\,\alpha^2\,u^2\,\text{cosh}\,(\Omega-h)$ and $\bar{Z}\,=\,\alpha^4\,u^4$. Finally, the Hawking temperature and entropy density now read
\begin{align}\label{anisotemp}
T=-\frac{f'(u_h)e^{-\chi(u_h)/2}}{4\pi}=\frac{3-m^2\,V(\bar{X}_h,\bar{Z}_h)}{4\pi u_h}\,e^{-\chi(u_h)/2}
\end{align}
and $s=\frac{2\pi}{u_h^2}$, both evaluated at the event horizon $u=u_h$.

To solve the equations above, we need to impose additional boundary conditions. At the black hole horizon, one has that $f(u=u_h)=0$ and $h(u=u_h)=h_h$. Close to the UV boundary $u=0$, we impose that $f(u=0)=1$, $\chi(u=0)=1$ together with the asymptotic expansion, 
\begin{align}\label{UVexp}
h\,(u)\,=\,\mathcal{H}_0\,+\,\dots\,+\,\mathcal{H}_3\,u^3\,+\dots\,.
\end{align}
Following the holographic dictionary, $\mathcal{H}_0$ is identified as the source of the stress tensor operator $T_{xy}$ in the dual system and will be set to zero hereafter \emph{i.e.}, $\mathcal{H}_0=0$, $\mathcal{H}_3$ corresponds to its VEV $\langle T_{xy}\rangle$. Within this setup, the source of $T_{xy}$ is solely contributed by the mechanical strain deformations induced by the axion fields. After solving the equations of motion numerically, one can analyze the nonlinear response by looking at shear stress~\cite{Ji:2022ovs}
$$\sigma\,\equiv\,\langle T_{xy}\rangle\,=\,\frac{3}{2}\,\mathcal{H}_3$$ 
as a function of the external mechanical strain \eqref{strain}. 

\begin{figure}[H]
    \centering
    \includegraphics[width=0.7\textwidth]{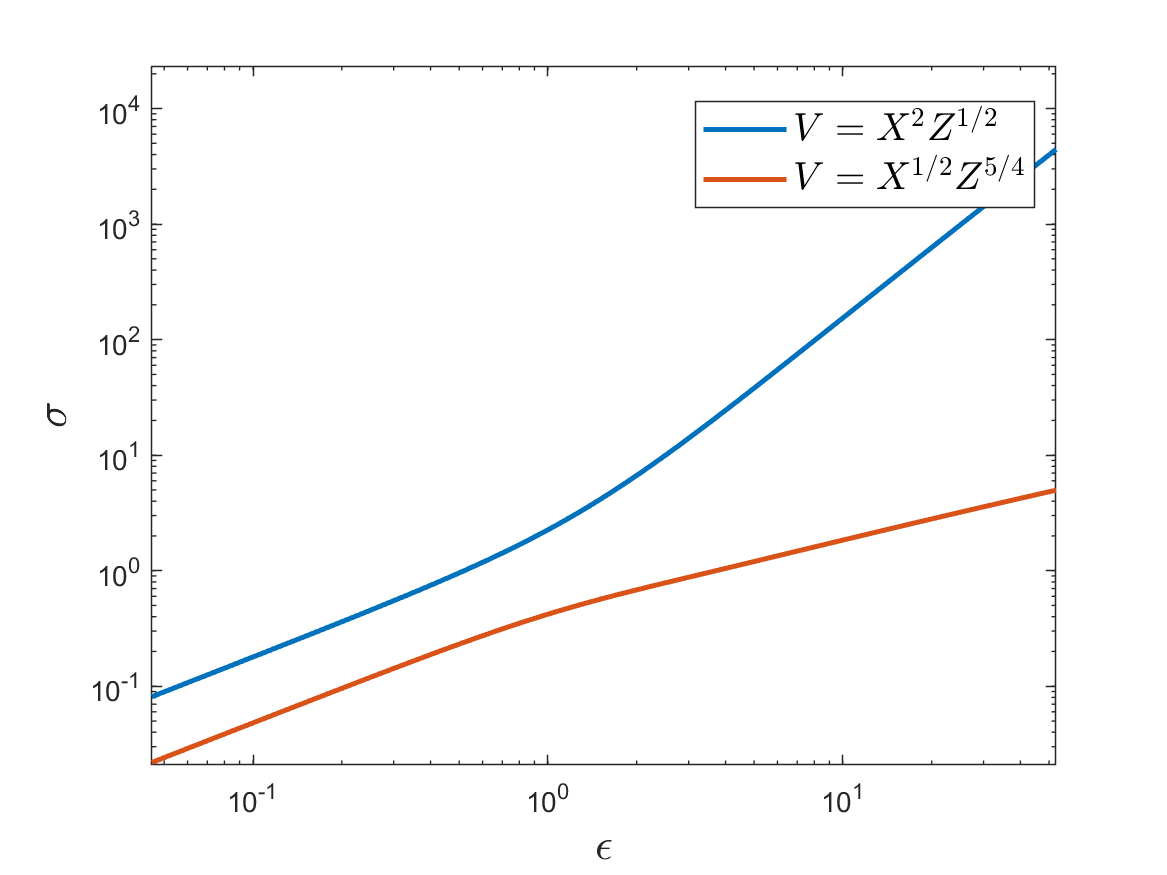}
    \caption{{The nonlinear elastic stress-strain curve for monotonic potential $V(X,Z)=X^NZ^M$. Here we choose $T/\alpha=0.2$. The blue curve exhibits shear hardening as evidenced by a rapid increase in the shear modulus $G\equiv d\sigma/d\epsilon$, while the red one corresponds to the shear softening cae characterized by the decrease of $G$ as the shear deformation is increased. For sufficiently large shear deformation $\epsilon$, a non-linear stress-strain scaling law of the form $\sigma\sim \epsilon^{3N/(2M+N)}$ emerges.}}\label{fig:example}
\end{figure}

In the case of monotonic potential $V(X,Z)$, the nonlinear elastic regime and its main properties (\textit{e.g.}, associated scaling behaviors) have been first explored in~\cite{Baggioli:2020qdg}. {For illustration, we show the stress-strain curves for shear softening (red) and hardening (blue) cases in Fig.~\ref{fig:example}. For the potential $V(X,Z)=X^NZ^M$, the non-linear stress-strain scaling law is given by $\sigma\sim \epsilon^{3N/(2M+N)}$ for large $\epsilon$.}
Furthermore, the thermodynamic and mechanical properties of strained holographic systems have also been computed and compared with the results from effective field theory and numerical simulations of amorphous solids in \cite{Pan:2021cux}, which unveiled some remarkable similarities between the two systems. The real-time nonlinear viscoelastic response has also been studied in~\cite{Baggioli:2019mck,Baggioli:2021tzr}. Finally, the DC thermoelectric transport properties of the normal phase and the dynamics of superfluid/supersolid phases under finite strain have been discussed respectively in~\cite{Ji:2022ovs} and~\cite{Baggioli:2022aft}. 

Despite all the aforementioned efforts, a systematic and complete stability analysis of the nonlinearly strained background is still missing. This will be the primary task of this work. 

\section{A simple holographic solid evading mechanical failure}\label{sec2}
In the following, we consider linear perturbations around backgrounds with a finite shear strain for different holographic models. As a consequence of the breaking of the $\text{SO}(2)$ rotational symmetry induced by the external shear strain, the dispersion relations of the low-energy modes now become angle-dependent. In order to capture this effect, we replace the ansatz \eqref{isolinear} for the fluctuating fields with the more general form,
\begin{align}\label{ansiolinear}
\delta \chi_A\sim \text{exp} \left(-\,i\,\omega\,t\,+\,i\,k\,x\,\text{cos}\,\theta\,+\,i\,k\,y\,\text{sin}\,\theta\,\right),
\end{align}
where $\theta$ denotes the angle between the wave vector $\Vec{k}$ and the $x$-axis. In this more complicated case, the fluctuations cannot be decomposed anymore in two decoupled longitudinal and transverse sectors and we need to solve all linearized equations at once. The resulting equations of motion for perturbations are obtained by substituting~\eqref{ansiolinear} into~\eqref{eom1}. They are too lengthy to be presented here\,\footnote{There are twelve coupled complex equations, ten of which come from the Einstein equations and two from the axion scalar sector.}. To compute the quasi-normal modes, we impose ingoing boundary conditions at the horizon and turn off all source terms at the AdS boundary.

Our numerical analysis indicates that, at least qualitatively, the dispersion relation of the hydrodynamic modes is not modified by the background strain. In particular, we verify the existence of two pairs of sound modes and one diffusive mode in the hydrodynamic regime (small $k$ limit) whose dispersions can be expressed as follows,
\begin{align}
    & \text{sound modes:}\quad\; \bold{\omega}\,=\,\pm\,v_{1,2}(\theta,\epsilon)\,k\,-\,\frac{i}{2}\,\Gamma_{1,2}(\theta,\epsilon)\,k^2\,,\label{smode}\\
    &\text{diffusive mode:}\;\;\bold{\omega}\,=\,-\,i\,D_\bold{\phi}(\theta,\epsilon)\,k^2\,.\label{dmode}
\end{align}
Here, we have denoted the sound mode with larger speed by the subscript ``1'' and the one with lower speed by the subscript ``2''. Following our notations, in the limit of $\epsilon \rightarrow 0$, the mode number ``$1$'' becomes the longitudinal sound mode, and the ``$2$'' the transverse one.  Finally, let us notice that all the hydrodynamic coefficients (sound speed, attenuation constant and diffusion constant) depend explicitly on $\theta$ and $\epsilon$. The same phenomenon appears in zero temperature field theory as well \cite{Alberte:2018doe}. Clearly, in the limit of $\epsilon \rightarrow 0$, the angle dependence disappears and the dispersion relations recover the well-know results presented in the previous section, Eqs.~\eqref{tmode}-\eqref{lmode}. 
\begin{figure}[t]
    \centering
       \includegraphics[width=0.32\textwidth]{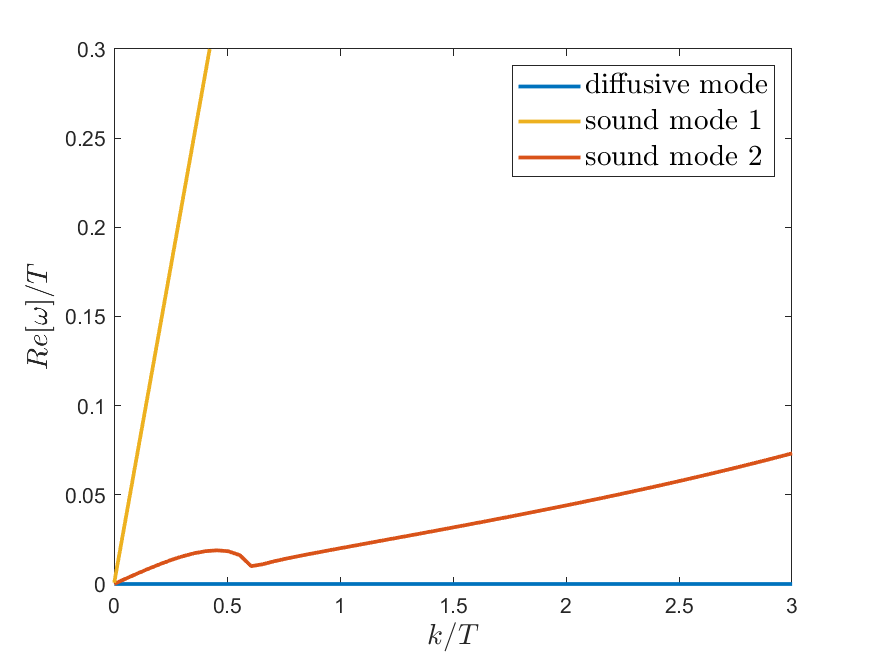}
          \includegraphics[width=0.32\textwidth]{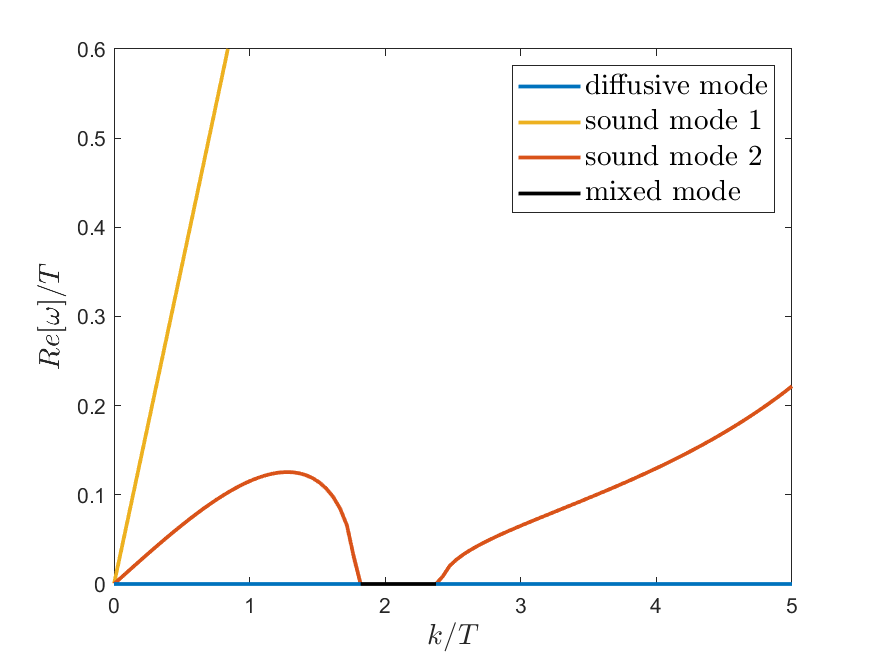}
             \includegraphics[width=0.32\textwidth]{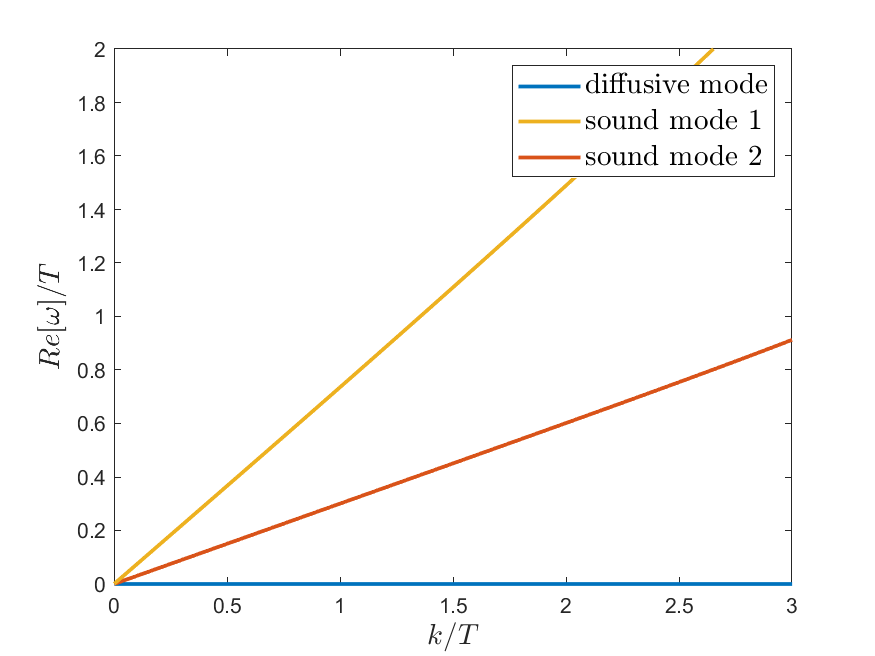}\\

                \includegraphics[width=0.32\textwidth]{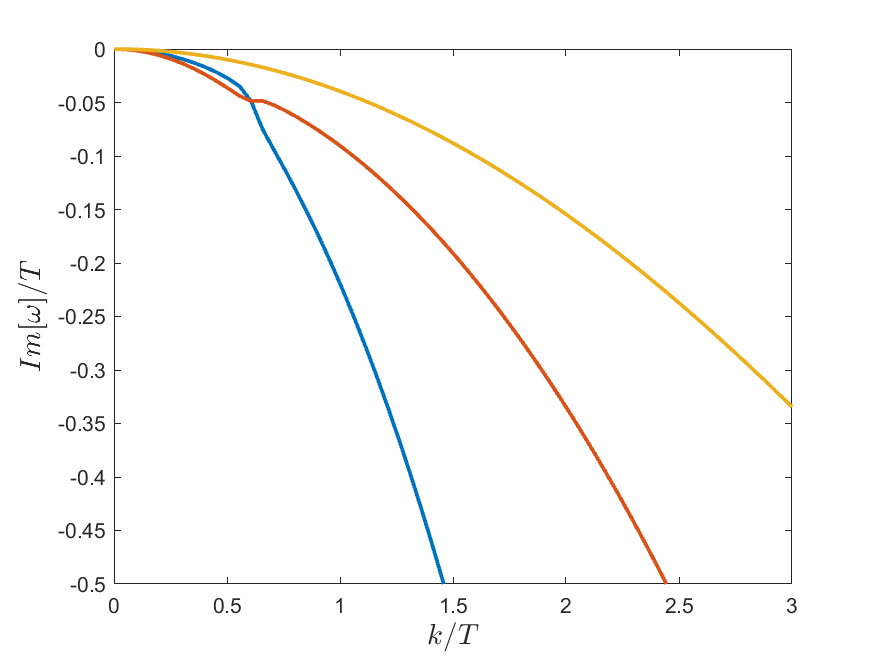}
                   \includegraphics[width=0.32\textwidth]{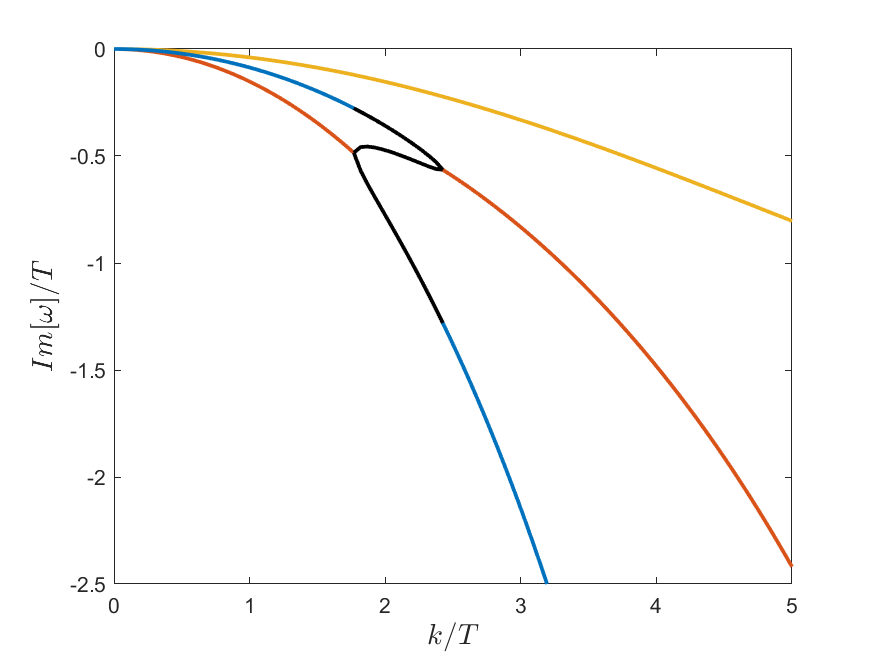}
                      \includegraphics[width=0.32\textwidth]{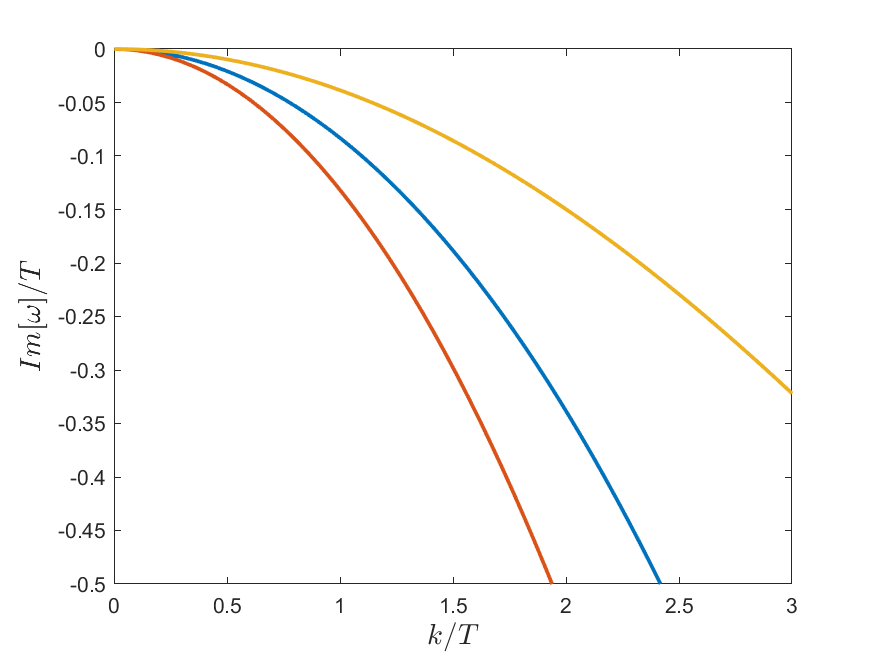}
    \caption{\textbf{Top panels.} The real part of the dispersion relations upon increasing the shear strain $\epsilon=1.28,2.05,3.01$ from left to right in the simplest model $V(X)=X^3$. \textbf{Bottom panels.} The corresponding imaginary parts. Here, we have set $T/\alpha=0.8$ and $\theta=0$.}
    \label{fig:4}
\end{figure}

For the simplest model with $V(X)=X^3$, the dispersion relations of the low-lying modes along the $x$-direction (\textit{i.e.}, $\theta=0$) have been depicted in Fig.~\ref{fig:4}. For small values of the background shear strain $\epsilon$ (top left panel), the sound mode number ``1" and the diffusive mode remain almost unaffected. On the contrary, the sound mode number ``2" presents a strong softening at an intermediate value of the wave-vector. This softening becomes stronger and moves to larger $k$ by increasing the background strain $\epsilon$. The softening of a phononic mode is typically a precursor for a structural instability and has been investigated in many situations, see \textit{e.g.} \cite{PhysRevLett.91.135501,RevModPhys.84.945,PhysRevLett.105.245502,PhysRevB.85.235407,PhysRevB.76.064120,PhysRevB.89.184111,Wang_2016}. Notice that despite this softening mechanism, the low-$k$ regime of the sound dispersion remains linear, even if with modified sound speed. Furthermore, we observe a strong interaction between the sound mode number ``2'' and the diffusive mode in the region below the softening.

Beyond a critical value of the strain (see central panel in Fig.~\ref{fig:4}), the interactions between these two modes become very complex and give rise to a propagation gap in the real part of the sound mode number ``2''. The ``propagation gap'' is a known phenomenon in the context of liquids and has been experimentally observed in liquid argon \cite{PhysRevLett.50.974,DESCHEPPER198429}, neon and in molecular dynamic simulations of Lennard-Jones liquids \cite{PhysRevA.29.1602}. Interestingly, it arises because of the strong intereference between elastic forces and dissipative effects \cite{DESCHEPPER19851}, which is at least qualitatively similar to what we observe.

Finally, for very large values of the strain (right panels in Fig.~\ref{fig:4}), the propagation gap closes and we see that the dispersion relations greatly simplify and return to a shape similar to the case with zero strain. Importantly, we see that all the modes in the system (including the non-hydrodynamic ones which are not shown in Fig.~\ref{fig:4}) have negative imaginary parts independently of the strength of the background strain $\epsilon$. This implies that this simple HHS is stable upon linear perturbations even when subjected to a very large background shear. Notice that this does not guarantee the stability of the system beyond linear order.

\begin{figure}
    \centering
   \includegraphics[width=0.49\textwidth]{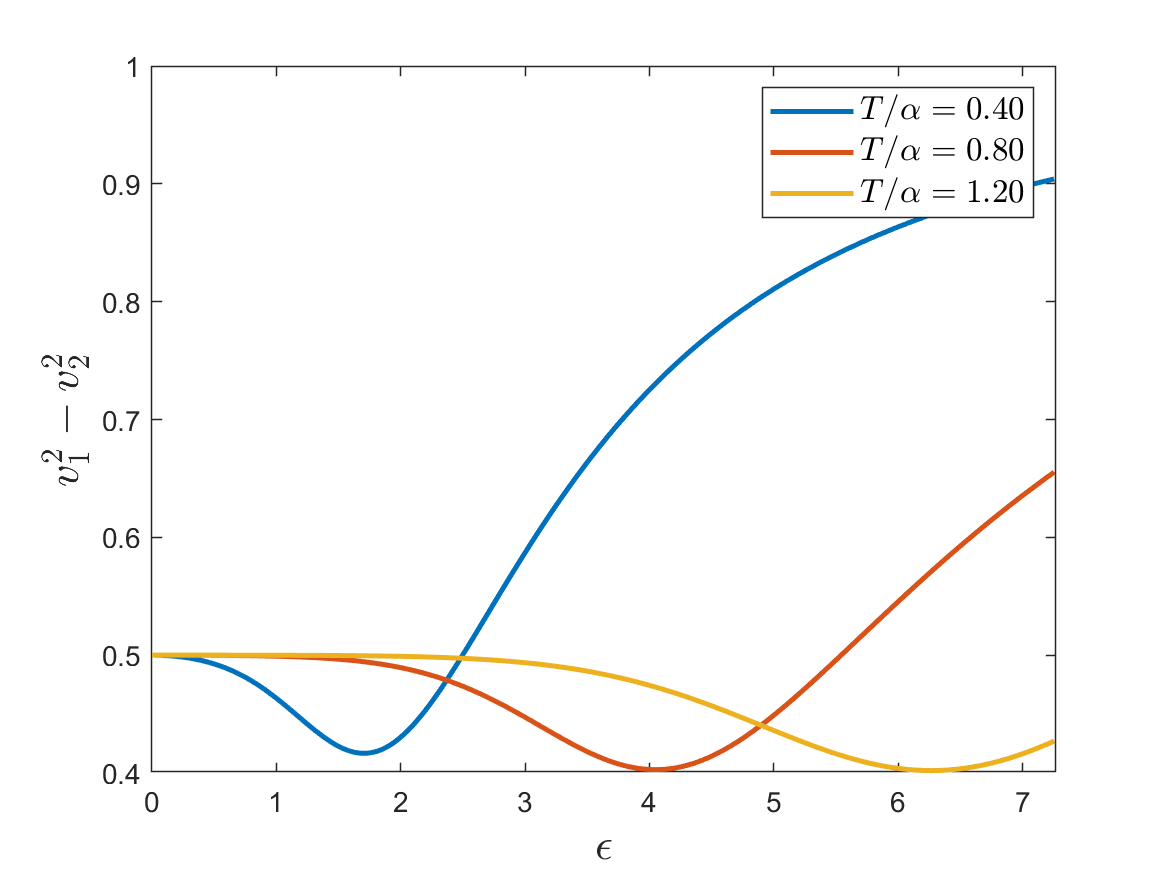}
    \includegraphics[width=0.50\textwidth]{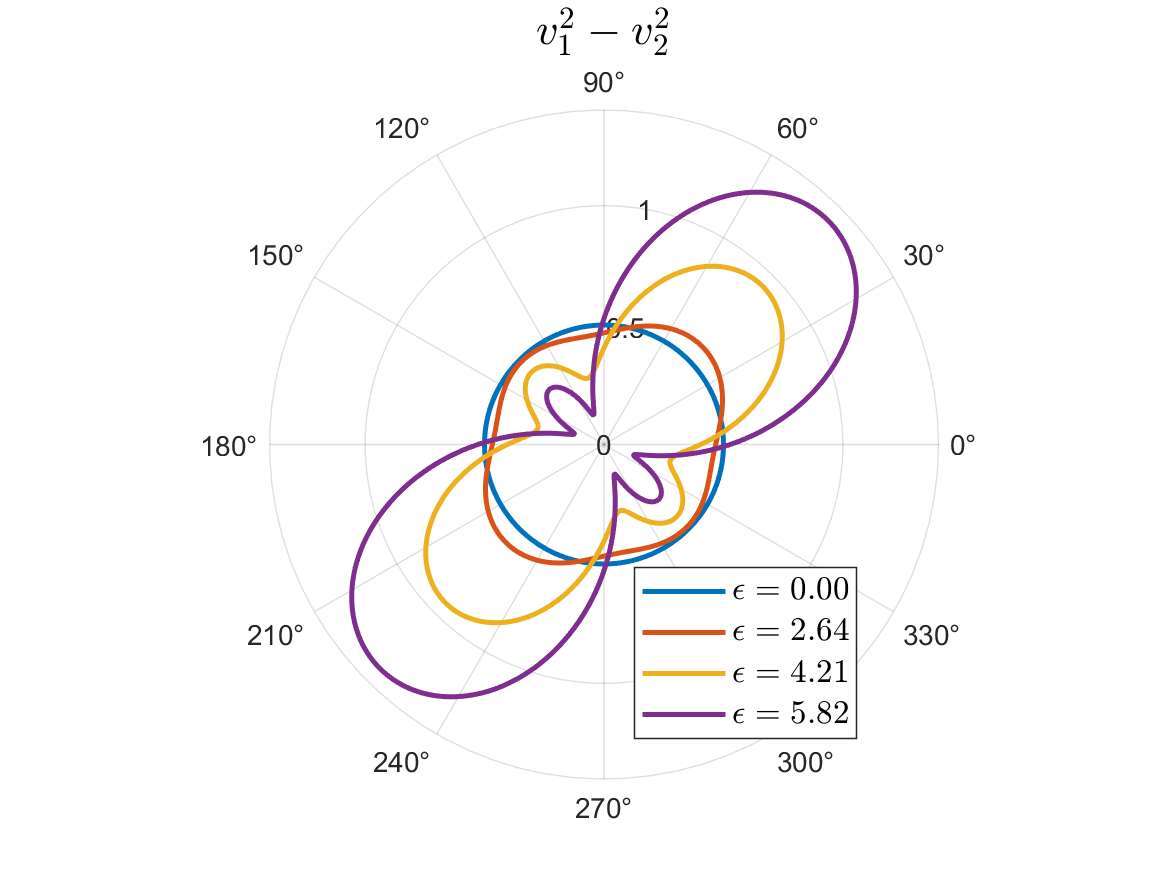}
    \caption{\textbf{Left: } $v_1^2-v_2^2$ as a function of the shear strain $\epsilon$ for $\theta=0$ and different values of the temperature $T/\alpha$. \textbf{Right: } the angular dependence of $v_1^2-v_2^2$ for different values of the background shear strain $\epsilon$ with $T/\alpha=0.8$.}
    \label{fig:8ccccc}
\end{figure}
In the long-wavelength limit (small $k$), we can compare our numerical results for different values of $\theta$ and $\epsilon$ with the dispersion relations presented in Eqs. \eqref{smode}-\eqref{dmode}. Doing so, we can extract all the hydrodynamic coefficients whose behavior is shown in Figs.~\ref{fig:7}-\ref{fig:8} of Appendix \ref{app1}. As anticipated, we find that
\begin{align}
v_{1,2}>0, \quad \Gamma_{1,2}>0,\quad  D_\phi>0 \label{positivecoe}
\end{align}
for all $\theta$'s and sufficiently large $\epsilon$.
A detailed analysis of the various coefficients as a function of the angle $\theta$ and $\epsilon$ can be found in Appendix \ref{app1}. Here, we limit ourselves to one single important observation. As a consequence of the conformal symmetry of our system, in absence of background strain, the two speeds of sound are related to each other by the following identity:
\begin{equation}\label{idi}
    {v_L}^2={v_T}^2+\frac{1}{2}\,, 
\end{equation}
which has been derived in \cite{Esposito:2017qpj} and confirmed numerically in several instances.
Here, we are interested in examining how this relation is modified when the velocities $v_{1,2}$ are considered and when the longitudinal and transverse excitations mix with each other.

The results are shown in Fig.~\ref{fig:8ccccc}. For simplicity, let us start from the behavior at $\theta=0$ (left panel). We observed that for small values of the shear strain $\epsilon$, the identity \eqref{idi} still holds. With the increase of the strain, the difference deviates from the $1/2$ value in a non-monotonic way. For very large strain, it becomes very anisotropic. As shown in Appendix~\ref{app1}, the effect of the background strain is to render the system anisotropic, by increasing one of the two speeds of sound and decreasing the other one. Furthermore, we observe that the effects of background strain are stronger for smaller temperatures. The larger the temperature, the larger the background strain which produces noticeable deviations from the $\epsilon=0$ state. Looking in more detail into the angular dependence of this difference (right panel of Fig.~\ref{fig:8ccccc}), {we observe that the anisotropy created by the background shear is of quadrupolar form corresponding to a discrete rotational invariance (see~\cite{Ji:2022ovs} for similar results regarding thermo-electric transport coefficients) and it is most pronounced in the diagonal direction $\theta=\pi/4$, $\theta=5\pi/4$. This is simply a consequence of the geometric structure of the shear deformation and it is independent of the strength of the latter. This discrete rotational symmetry, which also defines the principle axis of shear, is the only symmetry left in the problem since continuous rotational invariance is broken by the shear deformation.} Interestingly, for the class of models with potential of the form $V(X)=X^N$ ($N>3$), we find that they are all linear stable independently of the value of the background shear strain. One common feature for $V(X)=X^N$ from the stress-strain curve is that the shear modulus $G\sim \epsilon^3$ for large shear deformation.

\begin{figure}[htbp]
    \centering
    \includegraphics[width=0.48\textwidth]{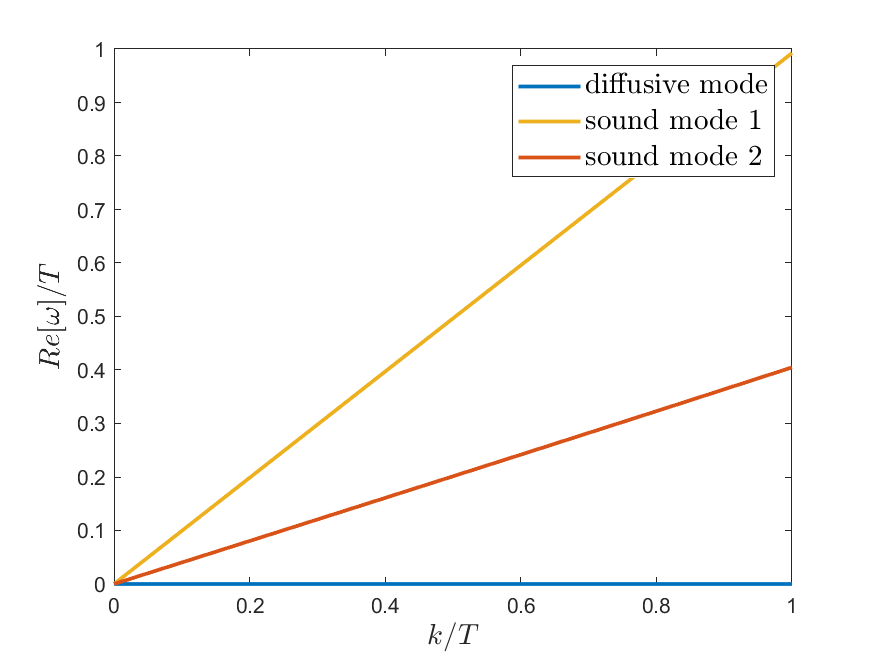}
    \includegraphics[width=0.48\textwidth]{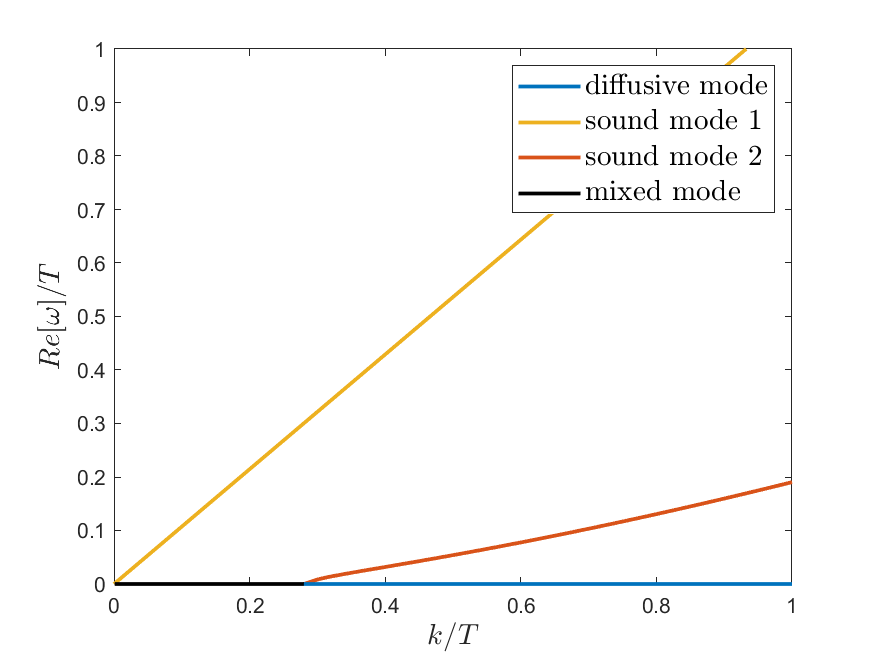}\\
     \includegraphics[width=0.48\textwidth]{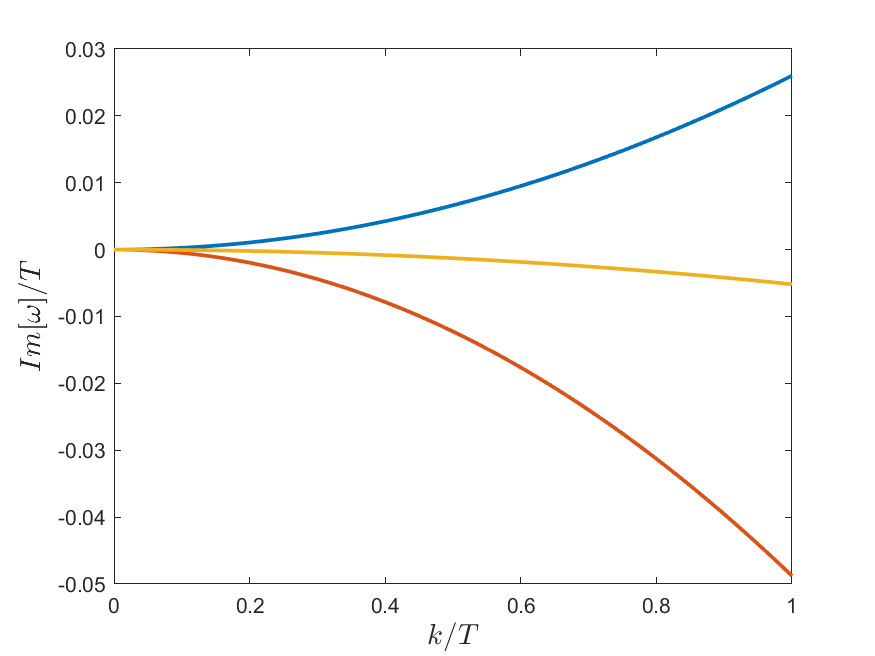}
    \includegraphics[width=0.48\textwidth]{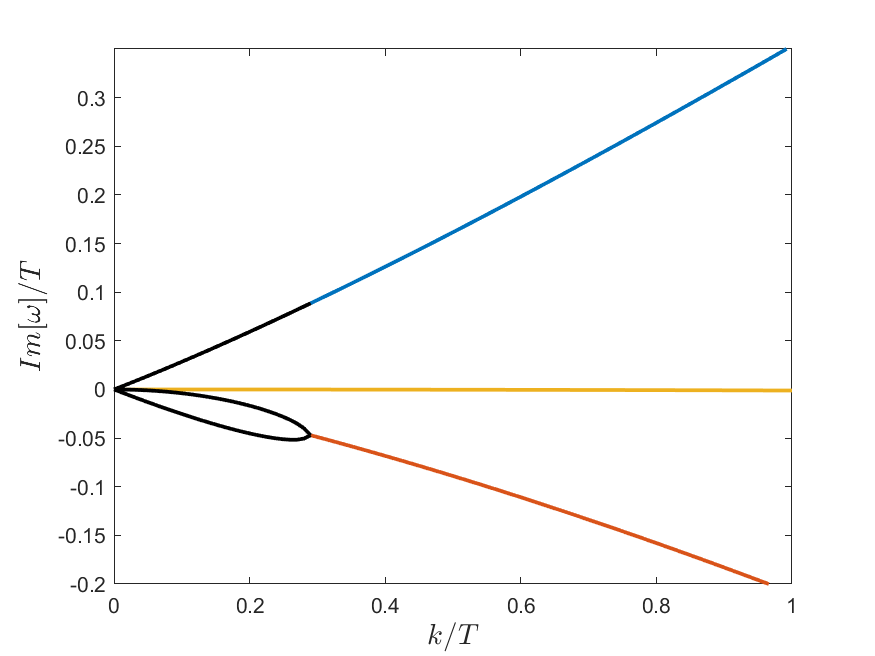}\\
    \caption{Dispersion relation curves for the $V(X,Z)=X^2\,Z^{1/2}$ model with $\epsilon=2$ (\textbf{left panels}) and $\epsilon=6.5$ (\textbf{right panels}). Here, we have fixed $T/\alpha=0.158$, $\theta=0$. It is found that critical values of the strain corresponding to a zero diffusion constant and a zero sound speed along the $x$-direction are given by $\epsilon_{0}^{\text{(d)}}(0)\approx1.75$ and $\epsilon_{0}^{\text{(s)}}(0)\approx5.55$.}
    \label{fig:X2Z_disp_diffusion_unstable}
\end{figure}

\section{A holographic solid exhibiting mechanical failure}\label{sec3}
In this section, we turn to considering HHS with a more complex potential of the form $V(X,Z)$ where an explicit dependence on the scalar quantity $Z$ is added. The main reason to do so is twofold. First, following an effective field theory logic, there is no argument to avoid such a dependence. Indeed, there is no symmetry principle which protects the potential not to depend on $Z$ (while there is one for having a $V(Z)$ without $X$ dependence, \textit{i.e.}, volume preserving diffeomorphisms). On the other hand, from a more phenomenological perspective, it is necessary to add such a dependence to encompass a broader spectrum of stress-strain curves including both strain hardening and softening behaviors.

Explicitly, we will consider the benchmark model $V(X,Z)=X^2Z^{1/2}$. We will comment at the end about another choice which gives qualitatively similar results. We have performed a detailed analysis of the quasi-normal mode spectrum as in the previous section. A complete report of such analysis can be found in Appendix \ref{app1}.
Differently from the previous simpler case, we have found that this setup exhibits two types of instability. The first type, which we denote ``diffusive instability'', is signaled by the diffusion constant $D_\phi$ becoming negative. And the corresponding critical strain is denoted as $\epsilon^{(d)}_0(\theta)$. An example of that sort is shown in the left panels of Fig.~\ref{fig:X2Z_disp_diffusion_unstable} for the case $\theta=0$. There, we observe the presence of a mode with dispersion:
\begin{equation}
    \omega_{\text{unstable}}=+i |D_\phi| k^2\,,
\end{equation}
which signals the appearance of a hydrodynamic instability for arbitrarily low values of the wave-vector $k$. As we will see in more detail later on, this type of instability is the dominant one for small values of the background shear strain $\epsilon$.\footnote{Note, however, that this instability was not observed in the previous effective field theory analysis at $T=0$ \cite{Alberte:2018doe}.} {We notice that the QNM spectrum presents five gapless modes as mentioned in~\eqref{smode} and~\eqref{dmode}. Each sound mode has two branches of dispersion relation that are mirror symmetric, \emph{i.e.} one $\Gamma$ corresponds to two real parts with $\pm v$, see ~\eqref{smode}. To avoid clutter, in Fig.~\ref{fig:4} and Fig. ~\ref{fig:X2Z_disp_diffusion_unstable}, we only show the $\mathrm{Re}[\omega]\ge 0$ branch. }
\begin{figure}[htbp]
    \centering
    \includegraphics[width=0.48\textwidth]{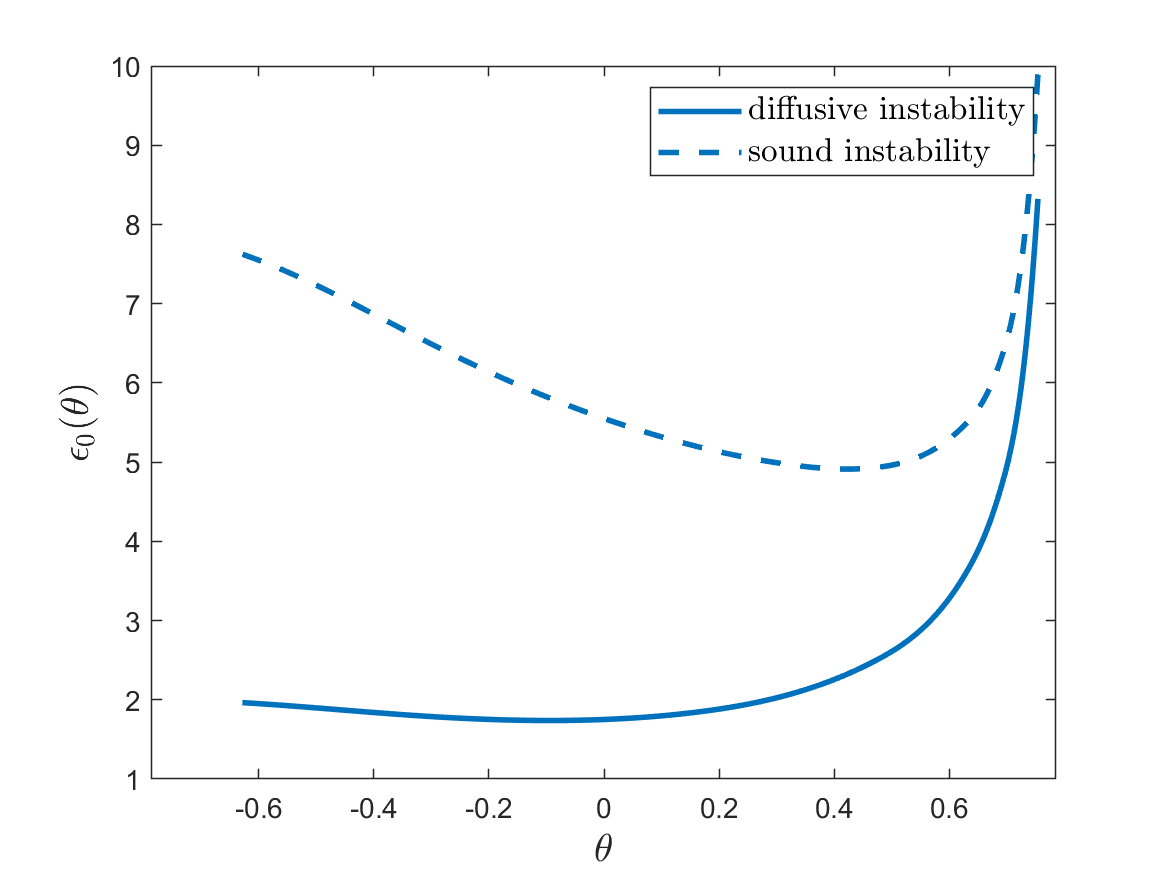}
    \includegraphics[width=0.48\textwidth]{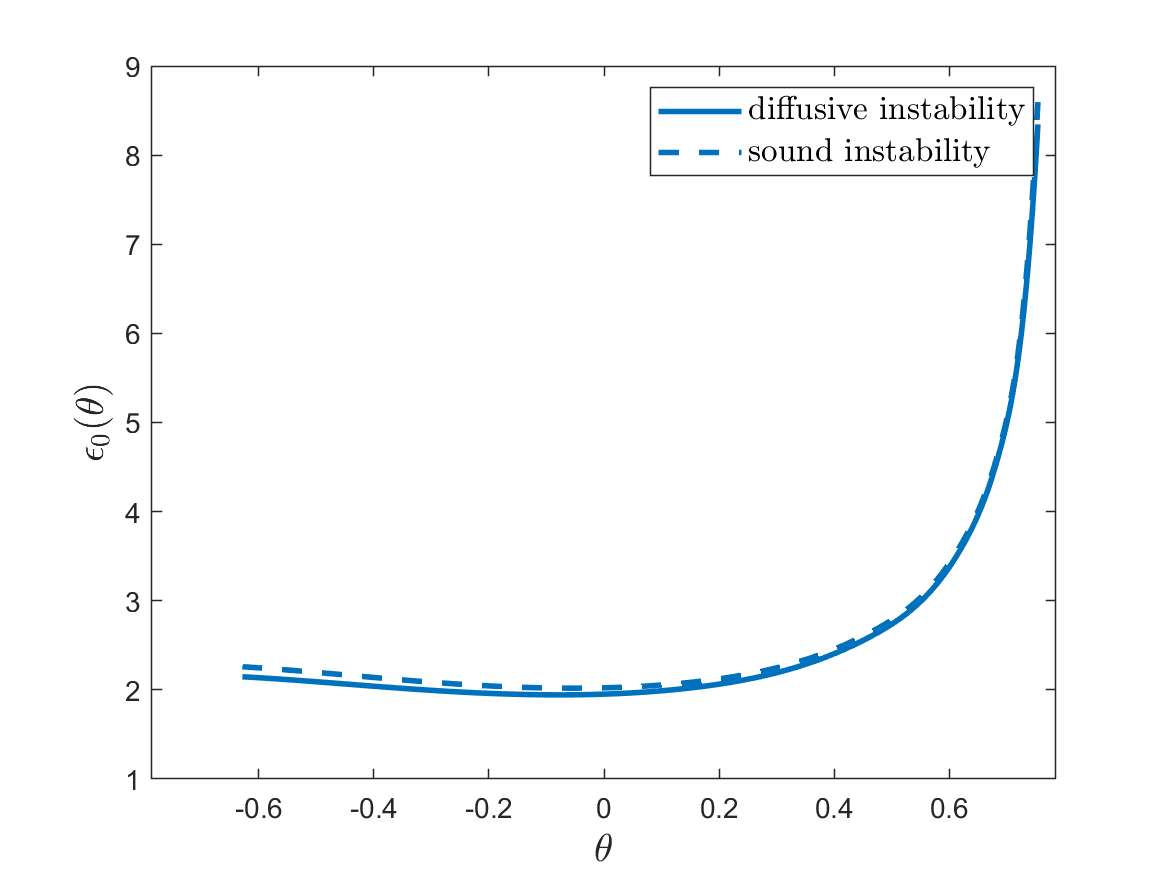}
    \caption{The angular dependence of $\epsilon_{0}^{\text{(d)}}$ and $\epsilon_{0}^{\text{(s)}}$ for $T/\alpha=0.158$ (left) and $T/\alpha=0.025$ (right). The critical strain that the system can sustain can be defined as the minimum of $\epsilon_{0}^{\text{(d)}}$.}
    \label{fig:X2Z_critical_angular}
\end{figure}


For larger values of the background shear strain, a second type of instability appears in the spectrum. We denote this second case ``sound instability'' since the dispersion of the unstable mode is given schematically by:
\begin{equation}
    \omega_{\text{unstable}}=+i |v|_{\text{unstable}} k - i \Gamma_{\text{unstable}} k^2.
\end{equation}
This second situation is shown in the right panels of Fig.~\ref{fig:X2Z_disp_diffusion_unstable} and it appears as the dominant instability for very large values of the background strain $\epsilon$. Intuitively, it corresponds to a speed of sound squared becoming negative. And we denote this critical strain as $\epsilon^{(s)}_0(\theta)$.

In Fig.~\ref{fig:X2Z_critical_angular}, we study in more detail the onset of these two instabilities by tracking the corresponding critical strain as a function of the angle $\theta$ for two characteristic temperature values, representing the low and high temperature regimes respectively. At high temperatures (left panel), we observe that the critical strain for the diffusive instability is always lower than that for the sound instability. The two approach each other only for large values of the angle, $\theta \rightarrow \pi/4$. Interestingly, for the leading diffusive instability the most unstable angle is $\theta\approx-\pi/36$. As shown in the right panel, the critical strain for the diffusive instability depends very mildly on the temperature. On the contrary, the one for the sound instability is strongly affected by thermal effects. In particular, by decreasing $T$ we see that the critical strain for the sound instability becomes smaller and approaches the one for the diffusive instability.

\begin{figure}[ht]
    \centering
    \includegraphics[width=0.48\linewidth]{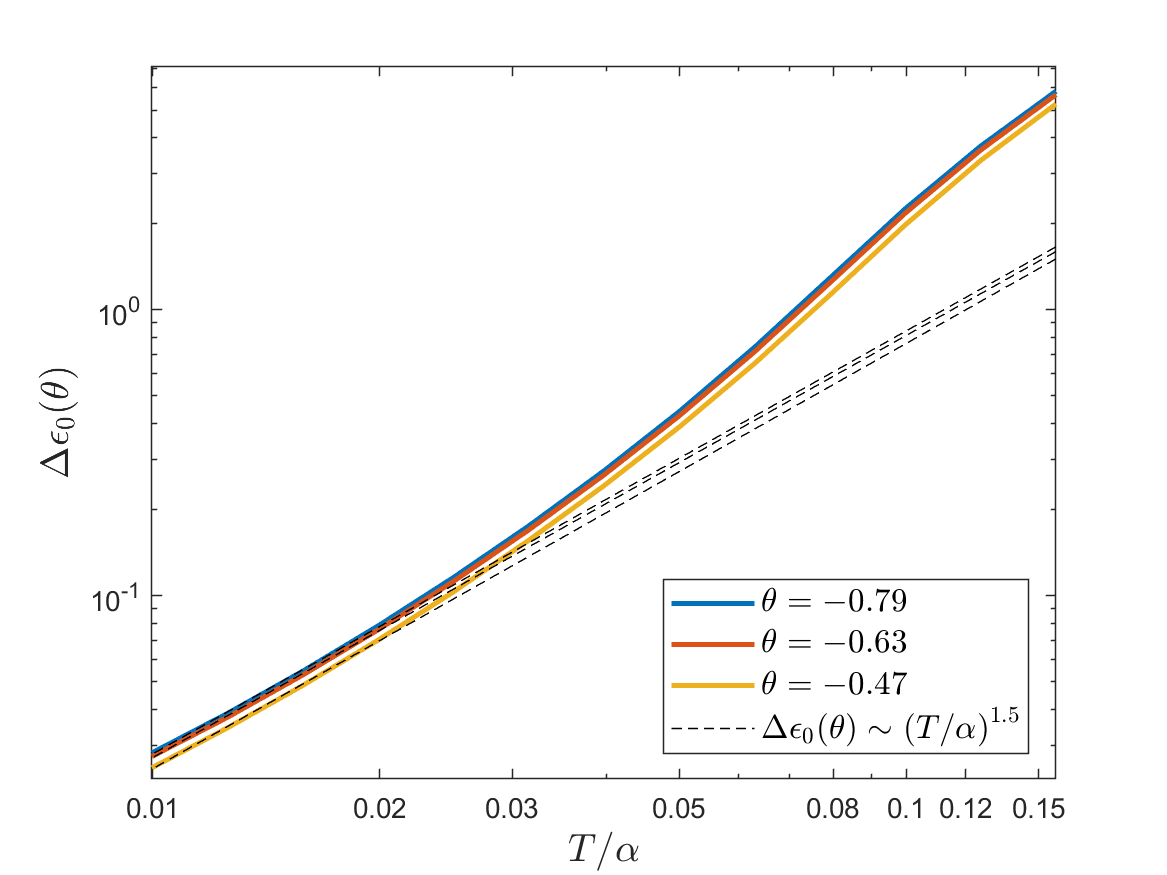} \quad \includegraphics[width=0.485\textwidth]{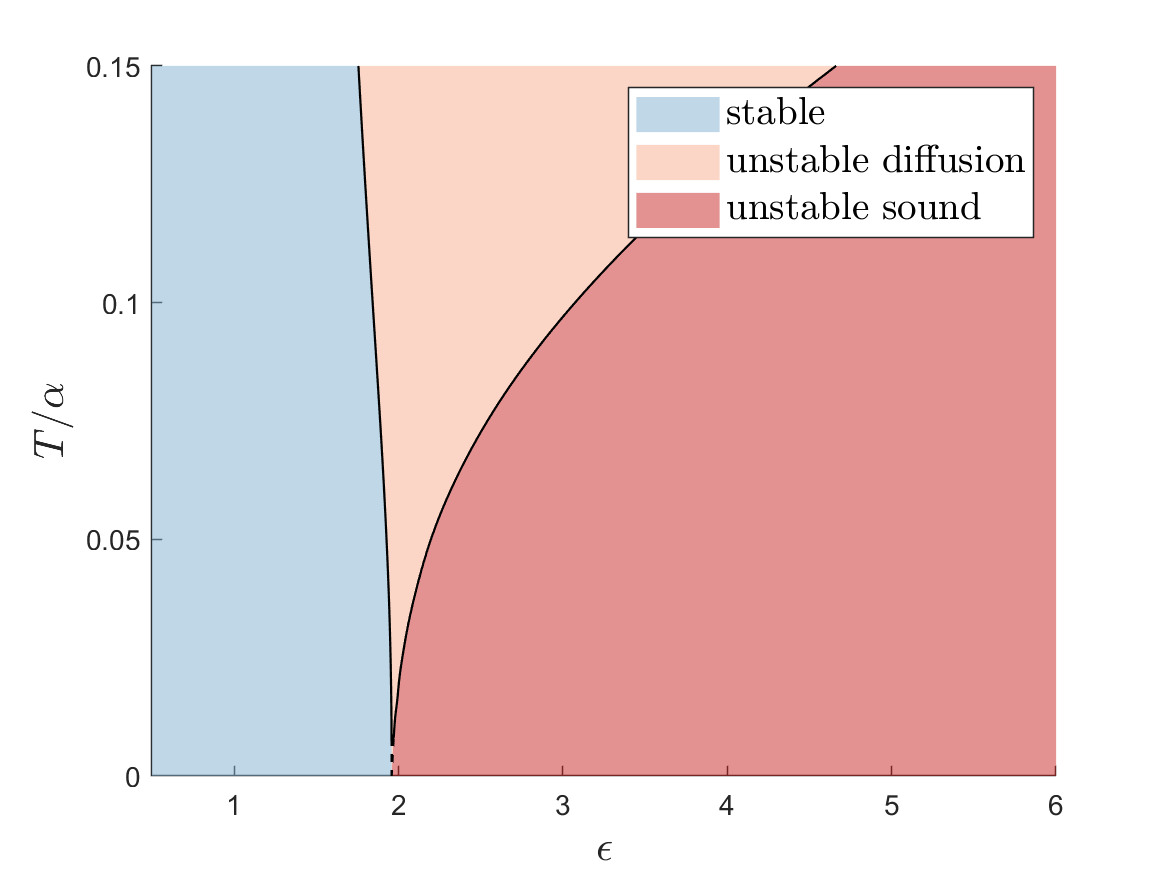}
    \caption{\textbf{Left: }The difference between the diffusive critical strain and the sound, $\Delta \epsilon_0\equiv \epsilon_0^{\text{(s)}}-\epsilon_0^{\text{(d)}}$ as a function of temperature for different angles $\theta$. The dashed lines illustrate the low temperature scaling $\Delta \epsilon_0 \propto (T/\alpha)^{3/2}$ as $T/\alpha\rightarrow0$. \textbf{Right: } The phase diagram as a function of temperature $T$ and background shear strain $\epsilon$ for the HHS model with $V=X^2 Z^{1/2}$. The instabilities are determined by the most unstable mode as a function of the angle $\theta$.}
    \label{fig:14}
\end{figure}

In order to investigate the temperature dependence of the critical strains, in the left panel of Fig.~\ref{fig:14} we show the difference between the critical strain corresponding to the sound instability, $\epsilon_0^{\text{(s)}}$, and that corresponding to the diffusive one, $\epsilon_0^{\text{(d)}}$, for different values of the angle $\theta$. There, we observe that independently of the angle $\theta$, the difference between the two critical values approaches zero following a power-law which is extracted numerically to be $\Delta \epsilon_0\propto T^{3/2}$.

We are now in the position to draw a phase diagram for our HHS at finite temperature $T$ and with background shear strain $\epsilon$. We show a representative result for the potential $V=X^2 Z^{1/2}$ in the right panel of Fig.~\ref{fig:14}. At low temperatures and low background strain (blue region), the holographic solid is stable and all excitations decay in time. By increasing the background strain $\epsilon$, the system becomes unstable since the diffusive mode moves in the upper half of the complex plane. This diffusive instability becomes more and more pronounced at large temperatures where dissipative effects are more important. As a consequence, the unstable diffusion region (pink color) is very thin at small $T$ and becomes larger and larger for higher $T$. By increasing the background shear strain $\epsilon$ further, the dominant instability is not anymore related to a diffusive mode but rather to a sound mode (red region). This instability is the dominant one at low temperatures and the only one surviving at $T=0$. By increasing the temperature, larger values of the strain are necessary to make this second instability the dominant one.

Interestingly, our phase diagram shares strong similarities with that obtained in \cite{Gouteraux:2022kpo} for a relativistic superfluid with background superflow and weakly broken translations (Fig.\,2 therein). The background shear strain $\epsilon$ plays the role of the superfluid velocity $\zeta$ as it similarly acts to destabilize the long-range order of the initial system. This analogy might suggest that also in our case the origin of the instability might be thermodynamics -- a thermodynamic susceptibility becomes negative at a critical value of the background strain. Qualitatively similar results are observed for $V(X,Z)=X^NZ^M$ with $5-2N-4M<0$ and $M\neq 0$. In contrast to the previous case with $M=0$, at large strain deformation, the stress-strain  relation now displays a rich power law behaviour $G\sim\epsilon^\nu$ with $\nu=\frac{3N}{2M+N}$. Therefore, our work shows that the same instability pattern emerges both in the shear softening $\nu<1$ and shear hardening $\nu>1$ cases.

\section{Outlook}\label{sec4}
In this work, we have performed a numerical study of the linear excitations in holographic homogeneous solids with background shear strain $\epsilon$. We have considered a large class of systems \cite{Baggioli:2021xuv} and analyzed in detail the behavior of the quasi-normal modes at finite temperature, zero charge density and finite $\epsilon$.
Surprisingly, we have found that the simplest HHS with potential of the form $V(X)=X^N$ ($N>3$) \cite{PhysRevLett.120.171602} are stable under linear perturbations independently of the value of the background shear strain. This suggests that instabilities under mechanical deformations can arise only at nonlinear level and would require a more sophisticated analysis which goes beyond the linearized equations of motion and quasi-normal modes.

On the contrary, for more complicated potentials of the form $V(X,Z)=X^N Z^M$ we have found a rich structure of instabilities which can be either diffusive or sound-like. The first type corresponds to a diffusion constant becoming negative; the second, to a speed of sound becoming complex. For small values of the shear strain $\epsilon$, we observed that the diffusive mode is the most unstable excitation and this becomes more prominent by increasing $T$, which naturally enhances dissipative effects and diffusive processes. On the other hand, for larger values of the shear strain, the dominant instability concerns a sound mode, as previously envisaged in the zero temperature field theory analyses \cite{Alberte:2018doe}.

Interestingly, the phase diagram constructed in this work (right panel of Fig.~\ref{fig:14}) shares intriguing similarities with that of a relativistic superfluid with weakly broken translations and background superflow \cite{Gouteraux:2022kpo}, where the superfluid velocity $\zeta$ plays the role of the shear strain $\epsilon$. It is feasible that also the instabilities revealed in this work have thermodynamic origin, and correspond to a thermodynamic susceptibility getting negative at a critical value of the strain $\epsilon$.

There are several open questions and point which deserve further attention and future investigation.
\begin{itemize}
    \item Our analysis has been mainly focused on the numerical study of the quasi-normal mode frequencies. In order to understand in more detail the origin of the revealed instabilities, it is necessary to write a hydrodynamic theory for systems with spontaneously broken translations under background shear strain. This would represent a generalization of the hydrodynamic framework reviewed in \cite{RevModPhys.95.011001} in presence of a background shear strain $\epsilon$ which breaks the underlying rotational symmetry. Such analysis would clarify whether the aforementioned instabilities have a thermodynamic origin as in the case of superfluids with superflow \cite{Gouteraux:2022kpo}.
    \item We have limited ourselves to the study of the linear dynamics. This does not allow to identify the endpoint of the instabilities and the resulting final equilibrium phase, which is expected to be inhomogeneous. In the same direction, we do expect that the linearly stable models $V(X)=X^N$ will develop nonlinear instabilities which are not visible in the linear regime. In this sense, the $X^N$ models could be driven to inhomogeneous phases by the background shear strain but only under finite, and eventually large perturbations. It is plausible that the unstable dynamics related to these models are similar to bubble nucleation processes \cite{ares2022effective,bea2022holographic,bea2021bubble,li2020holographic}, with a finite activation energy, while those related to the linearly unstable models (with either diffusive or sound instabilities) are more \textit{akin} to spinodal decomposition (see for example \cite{Attems:2019yqn,Attems:2020qkg,zhao2023dynamical}).
    \item A physical interpretation for these instabilities is still missing. In particular, the question whether these instabilities are related to any emergent plasticity or to the formation/dynamics of defects in these solids is still open. Here, we limit ourselves to notice that the same unstable dynamics are observed for holographic potentials corresponding to brittle and ductile stress-strain curves. This suggests that these instabilities do not depend on the UV microscopic physics (which is certainly very different for brittle and ductile materials) but only on some more universal IR features.
    \item In this work, we have limited our discussion to a benchmark potential of the form $V(X,Z)=X^N Z^M$. This gives a monotonic stress-strain curve that has a linear regime for small shear deformation and displays a power law behaviour $G\sim \epsilon^\nu$ with $\nu$ a constant for large strain deformations. It would be interesting to consider more complex mechanical responses, such as mixed effects of shear softening and shear hardening and a case with a ``yielding" point, characterized by a maximum on the stress-strain curve. { Moreover, our study can be extended to the case of three-dimensional materials by introducing three massless axions and extending the effective theory description appropriately. Despite we do not foresee fundamental difficulties, from a technical point of view the analysis is expected to be more cumbersome.}
\end{itemize}

\acknowledgments

We would like to thank Yuliang Jin, Deng Pan and Yun-Jiang Wang for helpful discussions. This work is supported by the National Natural Science Foundation of China (NSFC) under Grant No.12075298, No.12122513, No.12275038 and No.12047503. M.B. acknowledges the support of the Shanghai Municipal Science and Technology Major Project (Grant No.2019SHZDZX01) and the sponsorship from the Yangyang Development Fund. W.J.L is grateful for the financial support from the Fundamental Research Funds for the Central Universities (Grant No.DUT22LK07). W.J.L would also like to thank Institute of Theoretical Physics, Chinese Academy of Sciences for the nice hospitality and the sponsorship from the Peng Huanwu Visiting Professor Program in 2023.
\appendix

\section{Extended analysis}\label{app1}
It is manifest that the effect of the background strain is to render the system anisotropic. For completeness, in this appendix we show in more detail how sound speeds $v_{1,2}$, sound attenuation constants $\Gamma_{1,2}$ and diffusion constant $D_{\phi}$ in (\ref{smode}) and (\ref{dmode}) depend on the shear strain $\epsilon$ as well as the propagation angle $\theta$.
 
\subsection{Model 1}
\begin{figure}[h]
    \centering
    \includegraphics[width=0.49\linewidth]{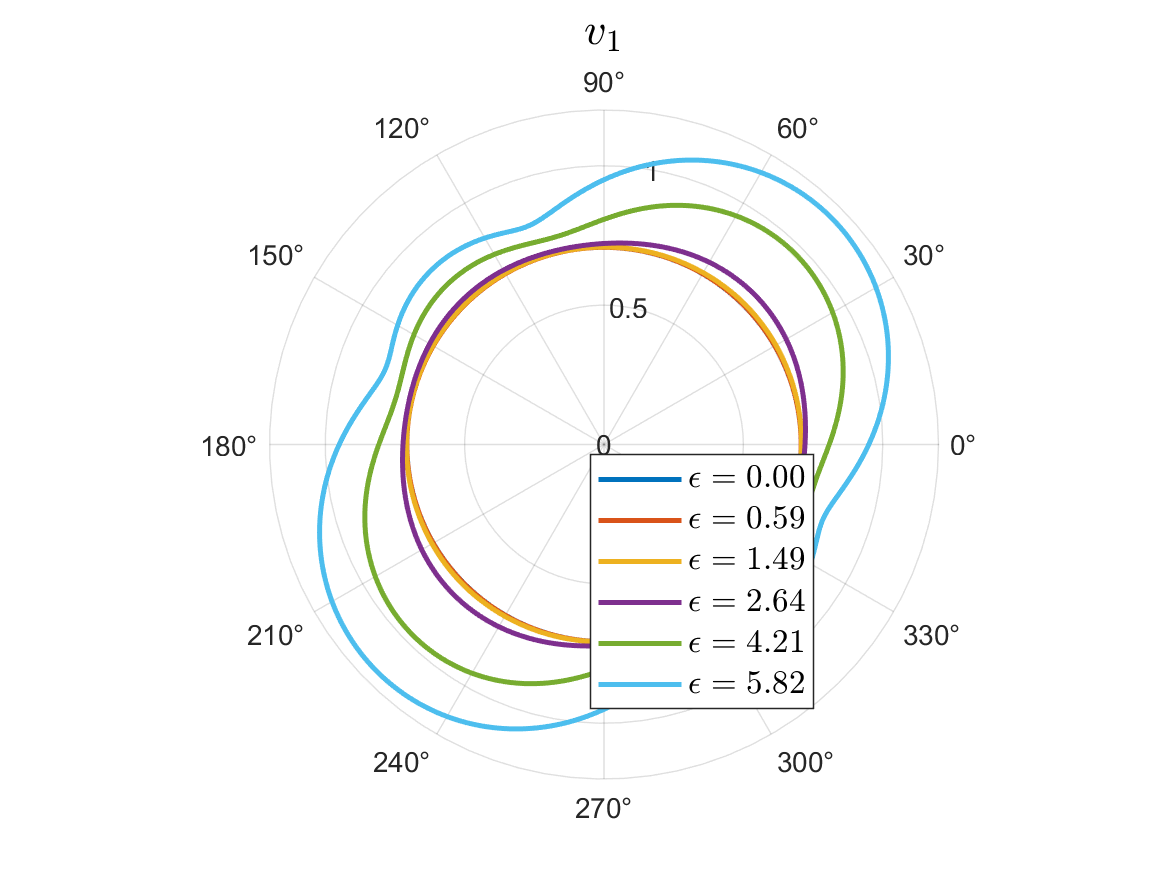}
    \includegraphics[width=0.49\linewidth]{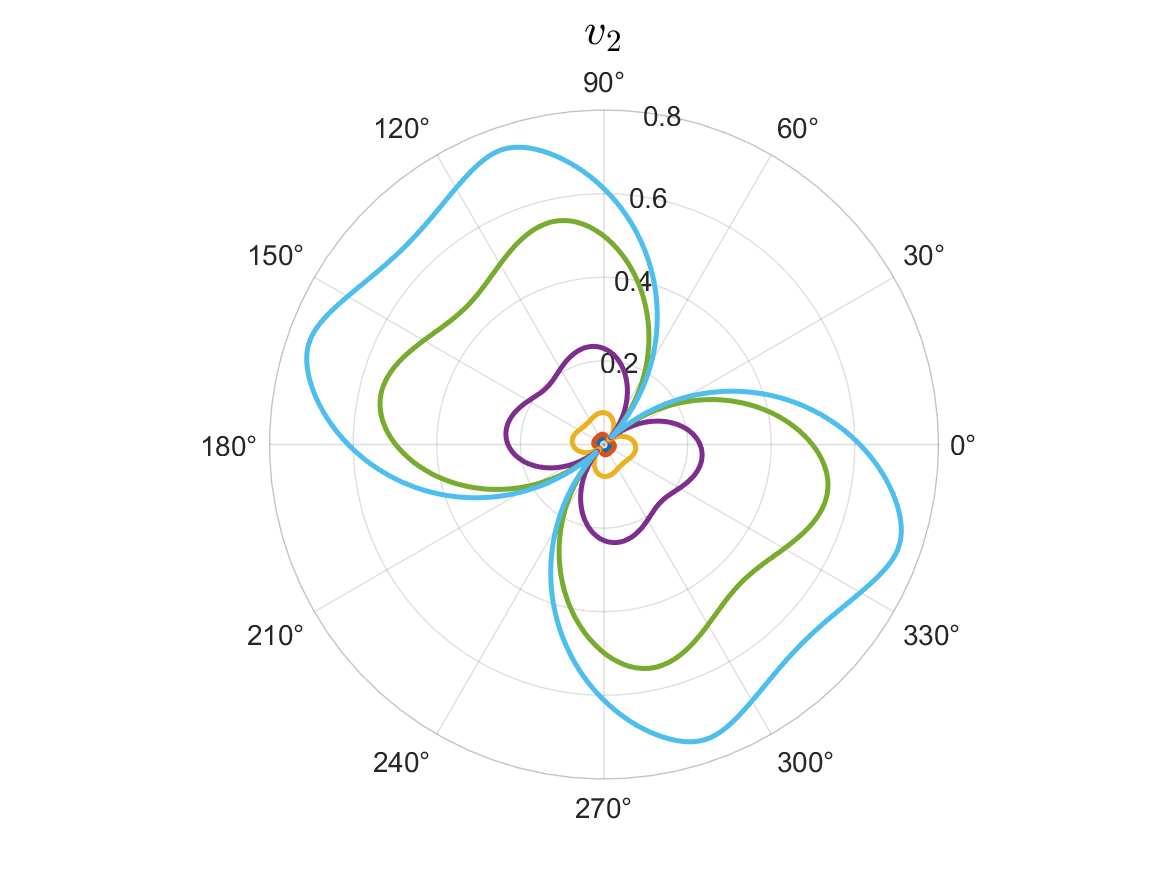}
    \includegraphics[width=0.49\linewidth]{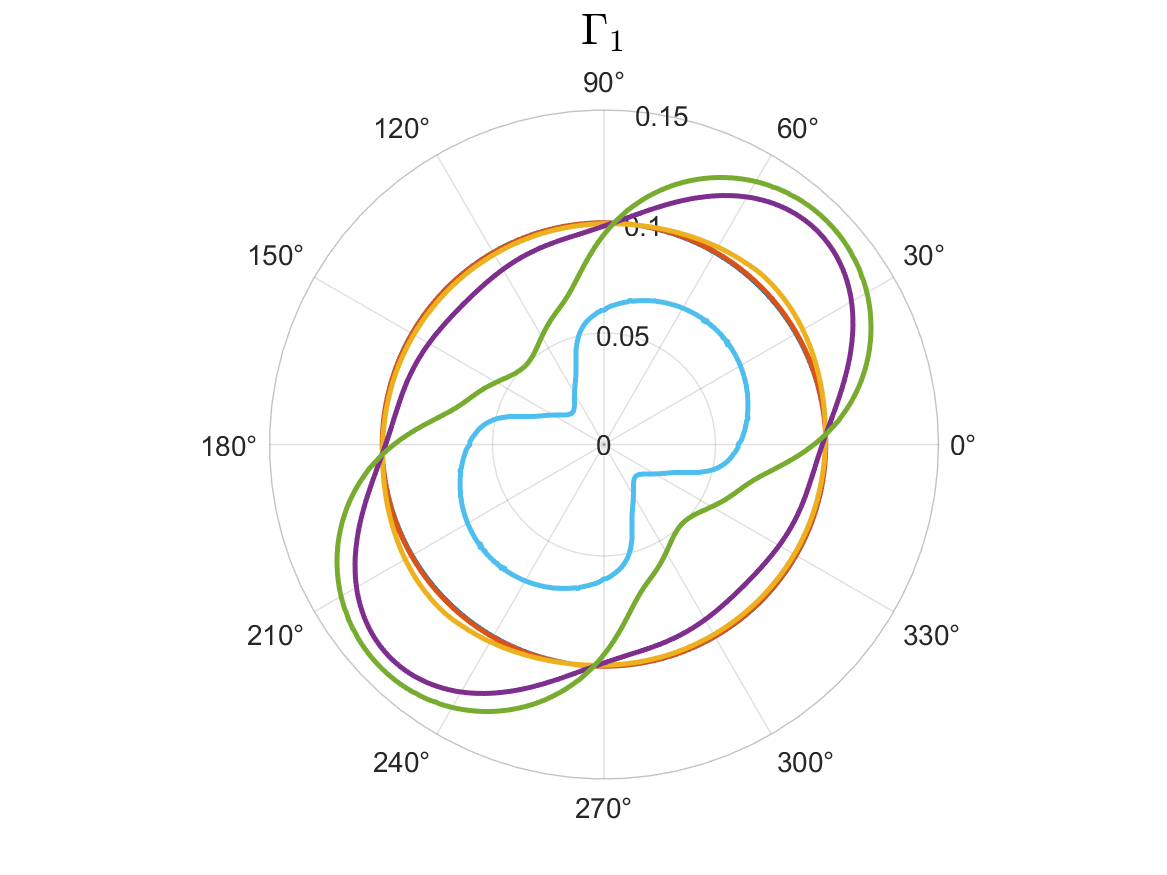}
    \includegraphics[width=0.49\linewidth]{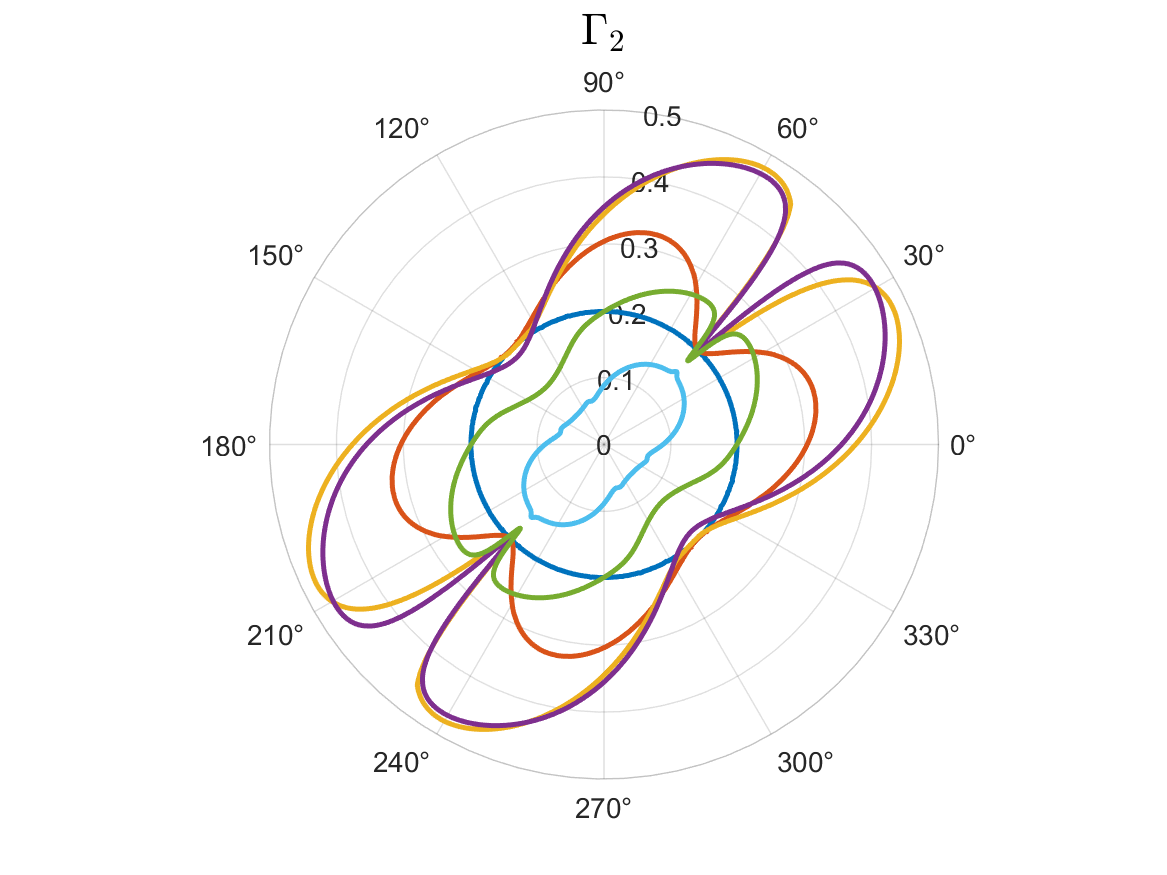}
    \caption{The simplest solid model with $V=X^3$ at $T/\alpha=0.8$: sound speeds $v_{1,2}$ (top panels) and sound attenuations $\Gamma_{1,2}$ (bottom panels) for  different values of strain and angle. In all cases, it is found that $v_{1,2}>0$ and $\Gamma_{1,2}>0$, which means that there is no sound instability.}
    \label{fig:7}
\end{figure}

\begin{figure}
    \centering
    \includegraphics[width=0.65\linewidth]{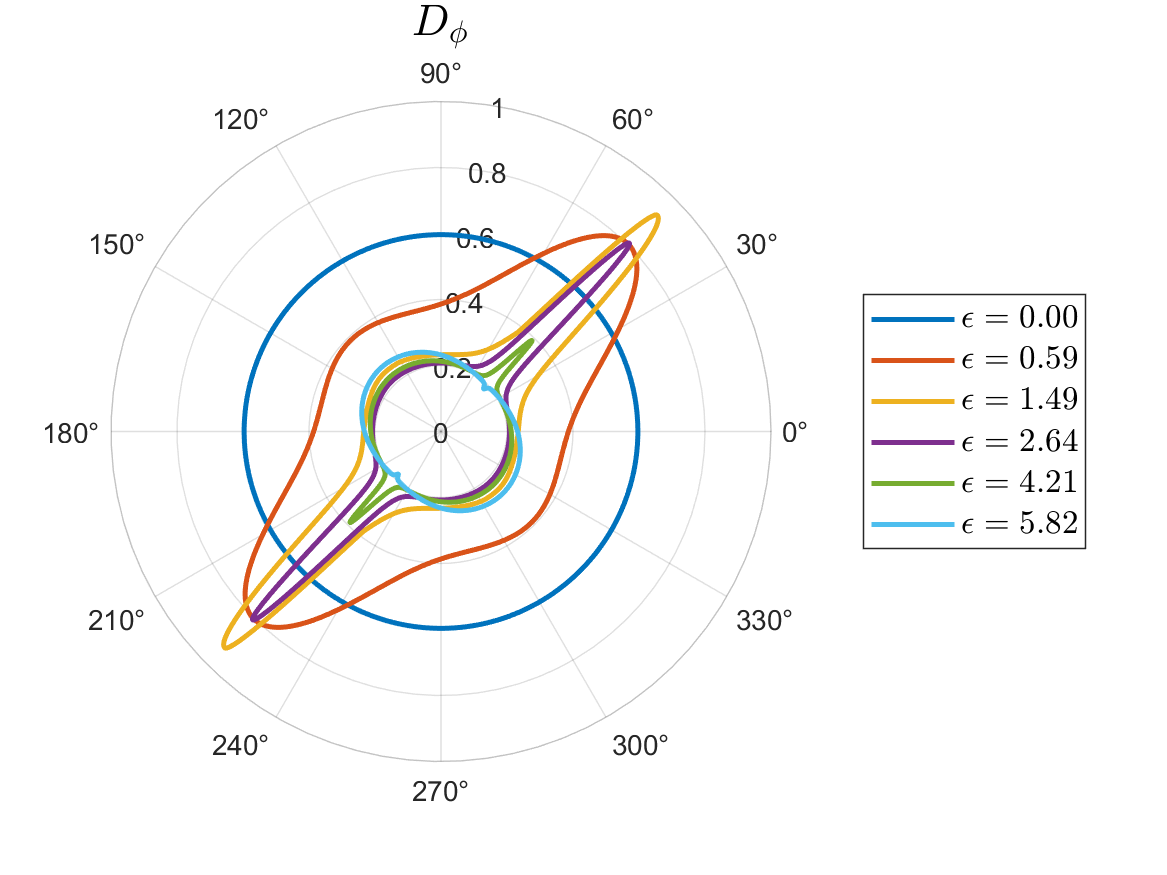}
    \caption{The simplest solid model with $V=X^3$ at $T/\alpha=0.8$: diffusion constant $D_\phi$ for  different values of strain and angle. In all cases, we find that $D_\phi>0$ which implies that there is no diffusive instability.}
    \label{fig:8}
\end{figure}

For the simplest model $V=X^3$ that evades mechanical failure, all the hydrodynamic coefficients as functions of $\theta$ and $\epsilon$ have been plotted in Figs.~\ref{fig:7}-\ref{fig:8} with a fixed $T/\alpha=0.8$. One finds that the anisotropy created by the background shear is of quadrupolar form and it is most pronounced in the diagonal direction $\theta=\pi/4$, $\theta=5\pi/4$. This could be simply a consequence of the geometric structure of the shear deformation. 

In Fig.~\ref{fig:7}, we find that $v_1$ is enhanced in all directions and $v_2$ is enhanced in most of the directions as the strain increases. More precisely, $v_1$ is enhanced most along the diagonal direction, \textit{i.e.}, $\theta=\pi/4,5\pi/4$. On the contrary, $v_2$ is only reduced close to the diagonal direction. The attenuation constants $\Gamma_1$ and $\Gamma_2$ depend on $\epsilon$ and the $\theta$ in a more complicated way. But anyway, we conclude that $v_{1,2}$ and $\Gamma_{1,2}$ are always positive, and therefore the system evades the sound instability no matter how large the strain is.  

For the diffusion mode, the behavior is similar to the attenuation constant $\Gamma_1$ (see Fig.~\ref{fig:8}). Importantly, the system also evades the diffusive instability.

\subsection{Model 2}
\begin{figure}
    \centering
    \includegraphics[width=0.48\linewidth]{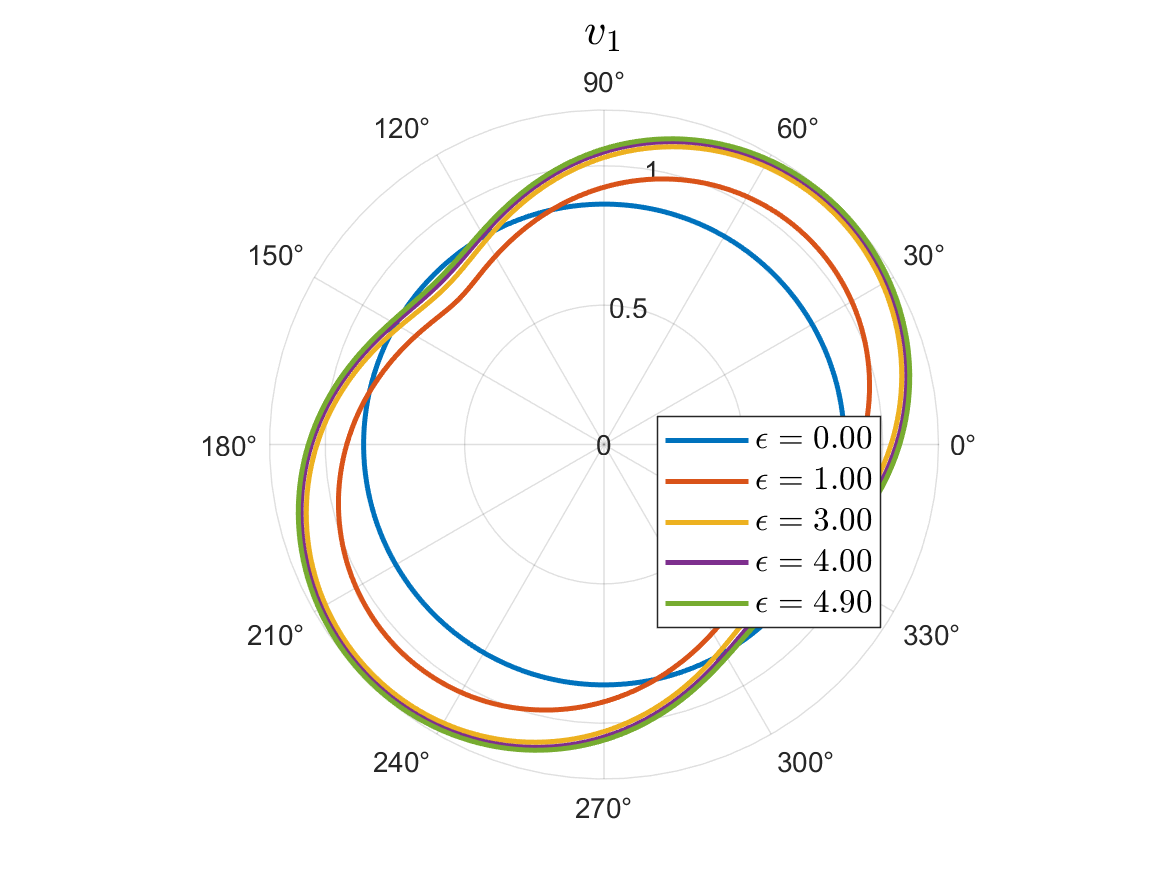}
    \includegraphics[width=0.48\linewidth]{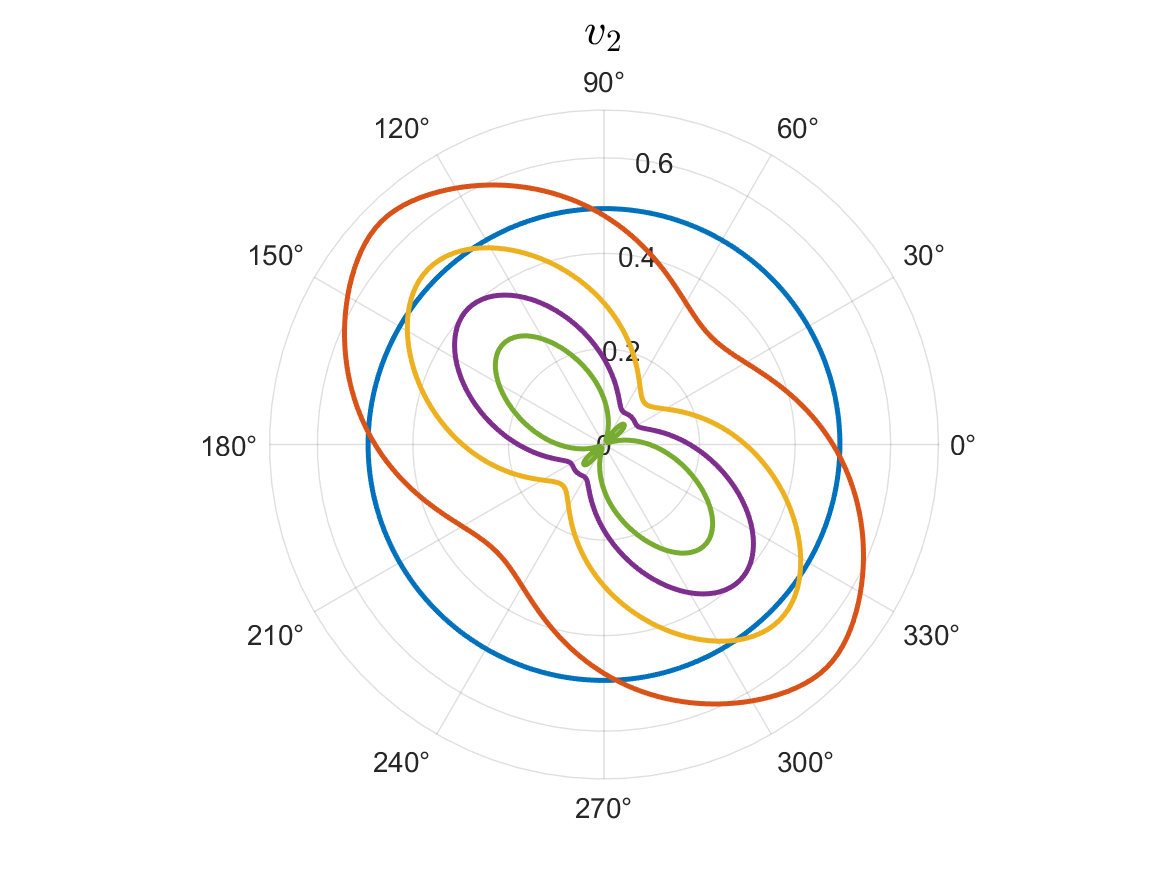}
    \includegraphics[width=0.48\linewidth]{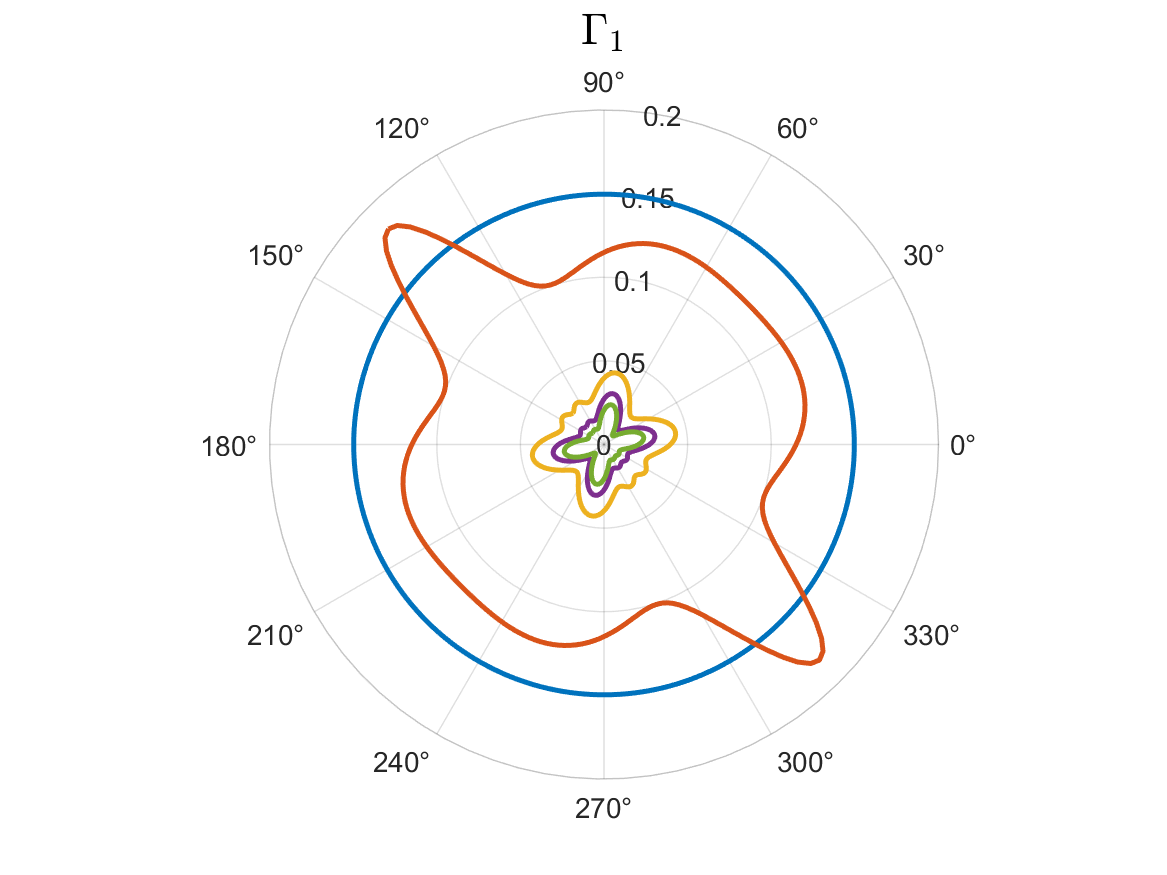}
    \includegraphics[width=0.48\linewidth]{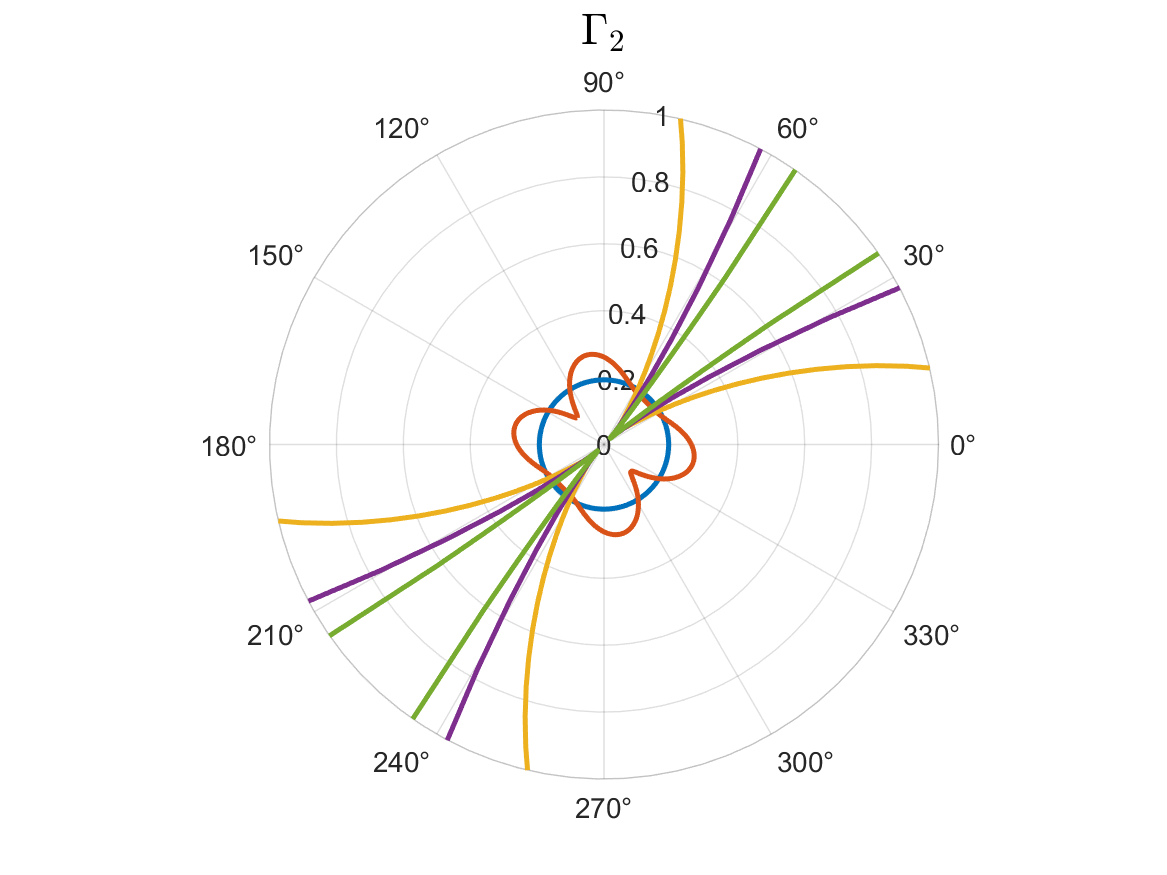}
    \caption{The solid model with $V=X^2Z^{1/2}$ at $T/\alpha=0.158$: sound speeds $v_{1,2}$ (top panels) and sound attenuations $\Gamma_{1,2}$ (bottom panels) for  different values of strain and angle. It is found that, when $\epsilon\approx 4.91$, the sound mode number ``2'' becomes unstable along $\theta\approx24^{\circ},66^{\circ},204^{\circ}$ and $246^{\circ}$ firstly.}
    \label{fig:11}
\end{figure}

\begin{figure}
    \centering
    \includegraphics[width=0.65\linewidth]{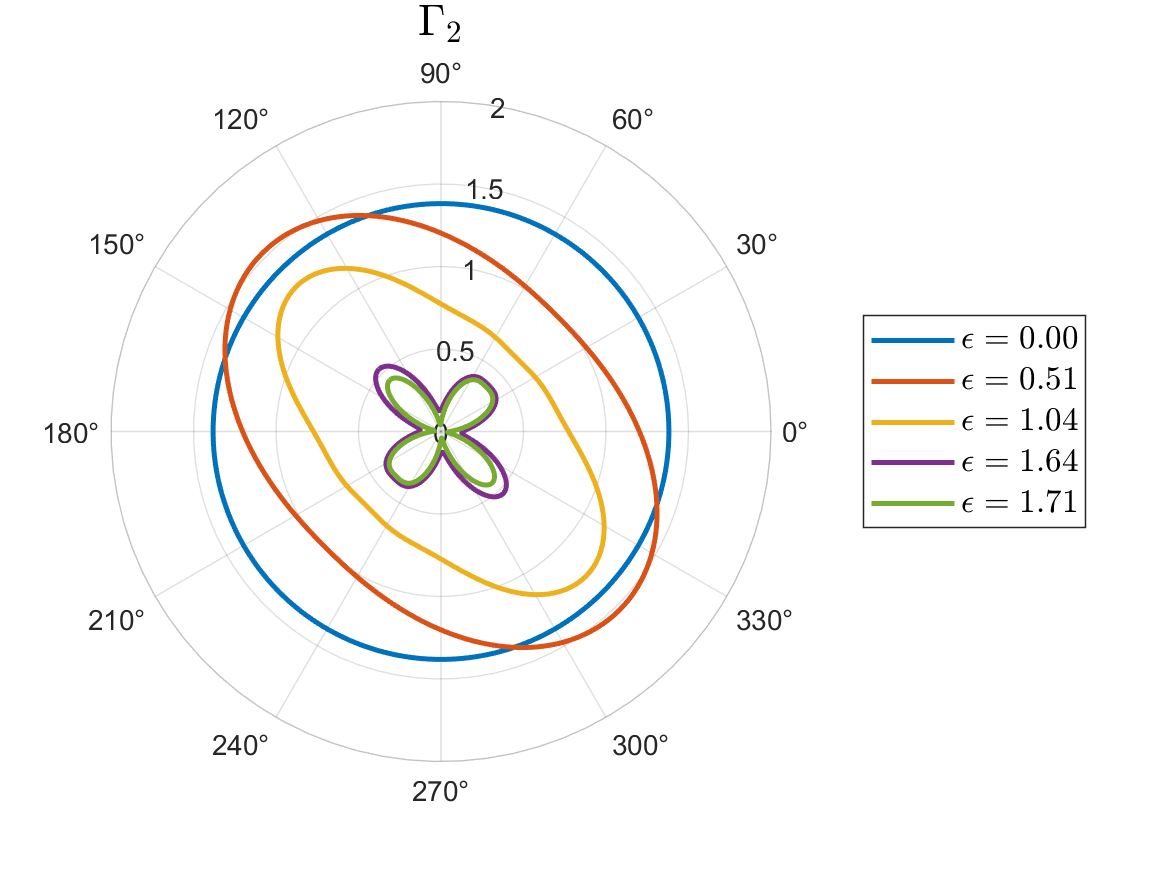}
    \caption{The solid model with $V=X^2Z^{1/2}$ at $T/\alpha=0.158$: diffusion constant $D_\phi$ for  different values of strain and angle. It is found that, when $\epsilon\approx1.74$, the diffusive mode becomes unstable along $\theta\approx95^{\circ},175^{\circ},275^{\circ}$ and $355^{\circ}$ firstly.}
    \label{fig:12}
\end{figure}

For the model with $V=X^2 Z^{1/2}$, all the hydrodynamic coefficients in this model as a function of $\theta$ and $\epsilon$ are presented in Figs.~\ref{fig:11}-\ref{fig:12} with a fixed $T/\alpha=0.158$.
Similar to the case above, the anisotropy has a quadrupolar form and is most pronounced along the principal axis due to the geometric structure of the shear deformation. Nevertheless, the angle dependence of those hydrodynamic coefficients are quite different from the previous model.

For the sound mode labelled as number ``1", we find that $v_1$ and $\Gamma_1$ are both positive in all cases. For the sound mode number ``2", it is found that, when $\epsilon\approx 4.91$, $v_2$ approaches to zero firstly along $\theta\approx24^{\circ},66^{\circ},204^{\circ}$ and $246^{\circ}$, suggesting the development of the sound instability. Whereas, $\Gamma_2$ always retains positive (even though it becomes very small when the applied strain is large). 

 The diffusion is strongly affected by the shear strain. When $\epsilon\approx 1.74$, the diffusion constant vanishes firstly along $\theta\approx95^{\circ},175^{\circ},275^{\circ}$ and $355^{\circ}$, which indicates the onset of the diffusive instability.

\clearpage
\bibliography{refs}
\bibliographystyle{JHEP}

\end{document}